\documentstyle[prl,aps,preprint,tighten,floats,aps,epsf,psfig]{revtex}

\def\abstract{\if@twocolumn
\section*{Abstract}
\else \normalsize
\begin{center}
{\bf Abstract\vspace{0pt}}
\end{center}
\setlength{\baselineskip}{4ex}
\quotation
\fi}
\def\endabstract{\if@twocolumn\else\endquotation\fi}

\begin{document}
\draft
\preprint{UPR-0793-T, hep-ph/9804428}
\date{\today}
\setcounter{footnote}{1}
\title{$U(1)^{'}$ Symmetry Breaking in Supersymmetric $E_{6}$ Models}
\author{Paul Langacker and Jing Wang}
\address{
          Department of Physics and Astronomy\\
          University of Pennsylvania\\
          Philadelphia, PA  19104\\
}

\begin{titlepage}
\maketitle
\def\thepage {}        

\begin{abstract}

We study the electroweak and $U(1)^{'}$ symmetry breaking patterns in models
with the particle content of supersymmetric $E_{6}$, including standard model
singlets $S$ and exotic quarks $D,~\bar{D}$. Motivated by free fermionic
string models, we do not require $E_{6}$-type relations between Yukawa
couplings. In particular, we assume that baryon and lepton numbers are
conserved, so that the exotic quarks can be light. Gauge invariance allows
Yukawa interactions between $S$ and Higgs doublets, and between $S$ and the
exotic quarks, allowing radiative $U(1)^{'}$ symmetry breaking and the
generation of an effective $\mu$ parameter at the electroweak scale. For both
the $E_{6}$ $\psi$ and $\eta$ models, universal soft supersymmetry breaking
parameters and Yukawa universality at the high (string) scale do not yield
acceptable low energy phenomenology. Relaxing universality, we find solutions
with phenomenologically acceptable values of $M_{Z^{'}}$ and the $Z-Z^{'}$
mixing angle. In addition, by varying the $U(1)^{'}$ charge assignments due to
the mixing of $U(1)_{\chi}$ and $U(1)_{\psi}$ of $E_{6}$, it is possible to
have acceptable low energy phenomenology with universal boundary conditions.         

\end{abstract}
\end{titlepage}

\section{Introduction}
It was argued in \cite{s1} that for a class of string motivated models with
extra $U(1)$'s, the $U(1)'$ breaking should be at the electroweak scale, and
in \cite{s3} a general analysis was done for a model in which the two standard
model (SM) Higgs doublets couple to a single SM singlet $S$ which carries a non-trivial $U(1)'$ charge. The breaking of the $U(1)'$ was shown to be at the electroweak scale, with a certain amount of fine tuning of the soft supersymmetry breaking parameters at $M_{string}$ required to suppress the $Z-Z^{'}$ mixing. In addition, the $U(1)^{'}$ symmetry forbids an elementary $\mu$ term, but an effective electroweak scale $\mu$ term arises when $S$ acquires a vacuum expectation value (VEV). It was argued that the radiative $U(1)'$ breaking is most easily accomplished if there are large Yukawa couplings between the singlet $S$ and exotic particles.

More complicated models involving two or more $S$ fields with opposite signs
for their $U(1)^{'}$ charges may lead to $U(1)^{'}$ breaking either at the
electroweak scale or an intermediate scale, depending on the details of the
superpotential and the supersymmetry breaking terms \cite{s4}.   

One of the major obstacles to analyzing $U(1)'$ breaking in string models is
that compactifications which have the SM gauge group and the SM particle
content with three generations and two Higgs doublets usually have additional
gauge symmetries and many exotic particles. The physics of $U(1)'$ and
electroweak breaking can be mingled with other issues such as the decoupling of heavy particles, preserving gauge unification in the presence of exotics, the communication between hidden and observable sectors, breaking of the extra non-Abelian symmetries, etc., that are not well understood. This motivated us to study in more detail the low energy phenomenology of the radiative $U(1)'$ breaking in a simpler model with all the elements that are necessary for a realistic theory.

Such a model should have the SM gauge group plus at least one extra $U(1)'$, three ordinary families, two or more Higgs doublets ($H_{1}$, $H_{2}$), a SM singlet ($S$) with nontrivial $U(1)'$ charge, and exotics ($D_{i}$). Phenomenological constraints suggest that the latter should be non-chiral \cite{s4A}. The model must be anomaly free, and the $U(1)'$ quantum numbers must be such that $Q_{H_{1}}+Q_{H_{2}}+Q_{S}=0$, $Q_{D_{1}}+Q_{D_{2}}+Q_{S}=0$ to allow such couplings as $\hat{S}\hat{H_{1}}\cdot \hat{H_{2}}$ and $\hat{S}\hat{D_{1}} \cdot \hat{D_{2}}$ in the superpotential. In addition, the model should have gauge unification at a scale comparable to $M_{string}$.      

A simple model satisfying these constraints is the $E_{6}$ model. $E_{6}$ is one of the most promising grand unification (GUT) models. Another motivation for $E_{6}$ is that Calabi-Yau compactifications of the heterotic superstring model lead to the gauge group $E_{6}$ or its subgroups in the observable sector, and also include the standard model representations for the matter multiplets as well as additional exotic fields \cite{s5}\cite{s5A}\cite{s5B}\cite{s5C}. It is not our intention to assume a full grand unified supersymmetric $E_{6}$ model. Strictly speaking, we use the particle content of the $E_{6}$ model, which provides acceptable anomaly free $U(1)'$ quantum numbers. Our purpose is to study the electroweak breaking of the model in the presence of an additional $U(1)'$ symmetry. 
In the full $E_{6}$ $GUT$ model, the Yukawa couplings of these exotics would
be related by $E_{6}$ to the Higgs Yukawa couplings so that the exotics would
have to be superheavy to avoid too rapid proton decay \cite{s12} (the
doublet-triplet problem). In our string motivated model the exotics and Higgs
Yukawa couplings do not respect the $E_{6}$ symmetries. In particular, the
dangerous exotic couplings may be absent, so that the exotics can be light. We 
also assume Yukawa universality, i.e., that all of the non-zero Yukawa
couplings are equal at the string scale.

Bounds from direct searches at the Fermilab Tevatron ($p\bar{p} \rightarrow Z'
\rightarrow l^{+}l^{-}$) \cite{s6} and precision electroweak tests \cite{s7}
on the $Z'$ mass and the $Z-Z^{'}$ mixing are stringent. The lower limits on
the $Z'$ mass are model dependent, but are typically around $500~GeV$
\cite{s6}\cite{s7} except in the case of suppressed coupling to ordinary
particles, such as in leptophobic models \cite{s8}-\cite{s8A}. Similarly, limits on the mixing angle are around a few $\times 10^{-3}$.

The particle content of the $E_{6}$ model we consider includes three $E_{6}$
27-plets, each of which includes an ordinary family, two Higgs-type doublets,
two standard model singlets, and two exotic $SU(2)$-singlet quarks with charge
$\pm 1/3$. In addition, there can be any number of vector pairs of chiral
supermultiplets from one or more $27+27^{*}$ representations without
introducing anomalies. For simplicity and consistency with the desired gauge unification, we assume a single pair of Higgs-type doublets from a $27+27^{*}$. Thus, there are eight Higgs doublet candidates, six singlet candidates and three exotic quark pair candidates.  
We assume that only a subset of these play a direct role in SM and $U(1)^{'}$ symmetry breaking, based on simple (string-motivated) assumptions concerning the non-zero Yukawa terms in the superpotential and the running of soft mass squared parameters being dominated by the largest Yukawa couplings. 
These are two $SM$ doublets $H_{1}$ and $H_{2}$, one $SM$ singlet $S$, and
exotic quarks $D$ and $\bar{D}$. We assume that all of the soft mass squares
are positive at the Planck scale. The soft mass squares of $H_{2}$ and $S$ are
typically both driven negative at low energy due to the Yukawa couplings
$\hat{Q}\hat{u}^{c}\cdot \hat{H_{2}}$ and $\hat{S} \hat{D} \cdot
\hat{\bar{D}}$ in the superpotential, where $Q$ and $u^{c}$ represent ordinary
quarks and antiquarks. Therefore, the $U(1)'$ breaking as well as the
electroweak breaking are radiative. The coupling $\hat{S}\hat{H_{1}}\cdot
\hat{H_{2}}$ in the superpotential becomes an effective $\mu$ term when $S$
acquires a VEV. Kinetic mixing \cite{s8A} \cite{s8B} \cite{s8C} between $U(1)_{Y}$ and $U(1)'$ is necessarily present in the theory from field theoretic loops, and its effects are included in the analysis. 

In \cite{s3} it was shown that there are two general scenarios to obtain a
small $Z-Z^{'}$ mixing angle. In one case, in which the symmetry breaking is
driven by large trilinear soft supersymmetry breaking terms, the $Z^{'}$ mass
is comparable to $M_{Z}$. Such a scenario is only allowed experimentally if
the $Z^{'}$ couplings to ordinary fermions are small, such as in leptophobic
models \cite{s8}-\cite{s8A}. We obtain examples of a light $Z^{'}$ and small
mixings (in fact, $M_{Z^{'}}<M_{Z}$). However, the couplings are not
leptophobic \footnote{Kolda et al \cite{s8A} have argued that in a class of
$E_{6}$ models kinetic mixing can yield leptophobic $Z^{'}$ couplings. We
include kinetic mixing in our model, but in contrast to the model considered
in \cite{s8A} (which included additional fields from an extra 78), the kinetic
mixing is too small to yield leptophobic couplings.}, and these cases are
excluded. The other possibility is that $M_{Z^{'}}$ is large, e.g., near the
$TeV$ scale, but the electroweak scale (and $M_{Z}$) is small due to
approximate cancellations. The signs of the $U(1)^{'}$ charges in usual
($\psi$ and $\eta$) $E_{6}$ models are such that one cannot obtain this
cancellation (even in the presence of kinetic mixing, which could in principle
change the signs) for universal boundary conditions (universal soft
supersymmetry breaking parameters with Yukawa universality). 
 However, with non-universal soft mass square parameters, there exist viable though somewhat fine-tuned solutions for a heavy $Z'$ with small mixing angle. We further utilize the fact that $E_{6}$ has two orthogonal $U(1)^{'}$ symmetries, and the surviving $U(1)^{'}$ \footnote{We restrict our analysis to the case of one extra surviving $U(1)^{'}$.} is in principle a linear combination with charge $Q=Q_{\chi}\cos{\theta_{E_{6}}}+Q_{\psi}\sin{\theta_{E_{6}}}$, where $U(1)_{\chi}$ and $U(1)_{\psi}$ refer to the patterns $SO(10) \rightarrow SU(5)\times U(1)_{\chi}$ and $E_{6} \rightarrow SO(10)\times U(1)_{\psi}$, respectively. With this additional degree of freedom, we find there are interesting solutions.  

In section II, we give the particle content and quantum numbers of the $E_{6}$
model. We also describe the explicit superpotential, the scalar potential, the
minimization conditions and the two phenomenologically acceptable scenarios
for $Z^{'}$. A brief outline of the RGE analysis is given in section III. In
section IV, we show that the universal boundary conditions can result in a
light $Z'$ and small mixing angle, in agreement with the large trilinear
coupling solutions of \cite{s3}. In particular, we find that $M_{Z'}< M_{Z}$
for both the $\psi$ and $\eta$ models. Because the effect of kinetic mixing
here is too small to make $Z'$ leptophobic in the $\eta$ model, this scenario
is not phenomenologically acceptable. In section V, we show that by deviating
from the universal boundary conditions at $M_{string}$, we can obtain a large
VEV for the singlet $S$, so that $M_{Z'}>500~GeV$. We argue that certain
stringent cancellation conditions have to be satisfied for the large $S$
scenario, which in turn requires fine tuning of the soft supersymmetry
breaking parameters at high energy. In section VI, we introduce the mixing
angle $\theta_{E_{6}}$ between the two $E_{6}$ $U(1)'$s. We find that with universal boundary conditions there exist acceptable solutions for $M_{Z'}$ and the mixing angle for certain parameter ranges of $\theta_{E_{6}}$ and $A_{s}$ (the trilinear coupling constant) for a given gaugino mass.       

The summary and conclusions are given in section VII. The renormalization group equations used in the analysis are presented in Appendix A. \\ 

\section{$E_{6}$ model and electroweak symmetry breaking.}

Two additional $U(1)'$ symmetries can occur when $E_{6}$ is broken, 
\begin{equation}
E_{6} \rightarrow SO(10) \times U(1)_{\psi} \rightarrow SU(5)\times U(1)_{\chi} \times U(1)_{\psi}
\end{equation}
The decomposition of the fundamental $27$ representation under the $SU(5)\times U(1)_{\psi}$ subgroup is 
\begin{equation}
27_{L} \rightarrow (10, 1)_{L}+(5^{*},1)_{L}+(1,1)_{L}+(5,-2)_{L}+(5^{*}, -2)_{L}+(1,4)_{L}
\end{equation}
where the first and second quantities are the $SU(5)$ representation and
$\sqrt{24} Q_{\psi}$, respectively, and the subscript means that left-chiral
fields will be assigned to the multiplets. $(10, 1)_{L}+(5^{*},1)_{L}$
correspond to an ordinary SM family, $(1,1)_{L}$ and $(1,4)_{L}$ are SM
singlets and $(5,-2)_{L}+(5^{*}, -2)_{L}$ are exotic multiplets which form a
vector pair under the Standard Model gauge group. $(5,-2)_{L}$ consists of
$D_{L}$ and $h_{2}$, where $D_{L}$ is a color-triplet quark with charge
$-1/3$, and $h_{2}$ can be either a Higgs doublet  or an exotic lepton
doublet. Similarly, $(5^{*}, -2)_{L}$ has the exotic antiquark $\bar{D}_{L}$,
and $h_{1}$, which can be either a down-type Higgs or a lepton doublet. Table
1 lists the $U(1)_{\chi}$, $U(1)_{\psi}$ and $U(1)_{\eta}$ charges of the
fields grouped under $SU(5)$, where $U(1)_{\eta}$ is a particular linear combination of $U(1)_{\chi}$ and $U(1)_{\psi}$, 
\begin{equation}
U(1)_{\eta}=\sqrt{\frac{3}{8}}U(1)_{\chi}-\sqrt{\frac{5}{8}}U(1)_{\psi},
\end{equation}
which occurs in Calabi-Yau compactifications of the heterotic string model if $E_{6}$ leads directly to a rank 5 group \cite{s10} via the Wilson line (Hosotani) mechanism.

Three $27_{L}$-plets are needed to form the three-family structure for the standard model. Gauge unification can be restored without introducing anomalies by adding a single $27_{L}+27_{L}^{*}$ pair, assuming that only the Higgs-like doublet $h_{2}$ and its conjugate $h_{3}$, associated with the $(5,-2)_{L}$ (from $27_{L}$) and $(5^{*}, +2)_{L}$ (from $27_{L}^{*}$), remain light, while the other fields from the $27_{L}+27_{L}^{*}$ pair acquire superheavy masses and decouple. Therefore, the light matter supermultiplets are 
\begin{equation}
3\times 27_{L} + (27_{L}+27_{L}^{*})|_{h_{2}+h_{3}},
\end{equation}
where the notation indicates that only the $h_{2}+h_{3}$ remain from the
$27_{L}+27^{*}_{L}$. Again, we want to emphasis that it is not our purpose to
consider $E_{6}$ as a GUT model, but to use its particle content and charge assignments as a concrete example to study $U(1)'$ symmetry breaking. In this model, 

(i) The additional $U(1)'$ is naturally anomaly-free as a result of being embedded in a larger gauge group. 

(ii) Any of the three $h_{1}$'s and four $h_{2}$'s can be MSSM Higgs doublets. The $S_{L}^{0}$ can be the singlet $S$ which couples to $H_{1}$ and $H_{2}$ in the superpotential. 

(iii) There are exotic quarks (antiquarks) $D$'s ($\bar{D}$'s), which have the same $U(1)^{'}$ charge as $h_{2}$'s and $h_{1}$'s, therefore allowing the Yukawa couplings $SD\bar{D}$ in the superpotential. 

The left chiral superfields of the model with their $SU(3)_{c}$, $SU(2)_{L}$, $\sqrt{5/3}Q_{Y}$ and extra $U(1)^{'}$ quantum numbers are listed in Table 2, where
 $\hat{d}_{i}^{c}$ are the left-handed down-type antiquarks; $\hat{D}_{i}$ and $\hat{\bar{D}}_{i}$ are the exotic color triplets. The gauge couplings are $g_{3}$, $g_{2}$, $g_{1}\equiv \sqrt{5/3}g_{Y}$ and $g_{1^{'}}$ for $SU(3)_{c}$, $SU(2)_{L}$, $U(1)_{Y}$, and $U(1)'$ respectively. 

With our assumptions that the couplings in the superpotential conserve baryon and lepton number, the most general form of the trilinear superpotential allowed by gauge invariance is 
\begin{equation}
W=h_{Q}^{ijk}\hat{u}^{c}_{i} \hat{Q}_{j} \cdot \hat{H}_{2k} +
h_{d}^{ijk}\hat{d}^{c}_{i}\hat{Q}_{j} \cdot \hat{H}_{1k}+h_{s}^{ijk} \hat{S}_{i} \hat{H_{1j}} \cdot \hat{H}_{2k}+ h_{D}^{ijk}\hat{S}_{i} \hat{D}_{j} \hat{\bar{D}}_{k}, 
\end{equation}
in which $i$, $j$, $k$ are family indices. The large Yukawa couplings dominate
the running of soft scalar mass squares and cubic coefficients, so we only
take the Yukawa coupling\footnote{The $b$ quark Yukawa coupling could be large 
for large $\tan{\beta} \equiv \langle H_{2}^{0} \rangle / \langle H_{1}^{0}
\rangle$.} of the top quark, $h_{Q}^{333}\hat{u}^{c}_{3} \hat{Q}_{3}
H_{23}$. Similarly, we assume that only $h_{s}^{333} \hat{S}_{3} H_{13}
H_{23}$ and $h_{D}^{333}\hat{S}_{3} \hat{D}_{3} \hat{\bar{D}}_{3}$ are
significant. This physical ansatz is motivated by perturbative string models,
in which the Yukawa couplings are all zero or of order $g_{0}$ (the gauge
coupling at the string scale)\cite{s5A}\cite{s5B}\cite{s5C}. In our numerical
work, we assume Yukawa universality, i.e., the non-zero Yukawa couplings at
the string scale are all equal; we take the value $g_{0}$ \cite{s4}
for definiteness. In this case, the superpotential is 
\begin{equation}
W=h_{Q}\hat{u}^{c}_{3} \hat{Q}_{3} \cdot \hat{H_{2}}+h_{s} \hat{S} \hat{H_{1}} \cdot \hat{H_{2}}+h_{D} \hat{S} \hat{D} \hat{\bar{D}}.
\end{equation}
where family indices have been suppressed. We assume the soft supersymmetry breaking terms 
\begin{eqnarray}
-L_{SB} & = & (\sum_{i}{M_{i}\lambda_{i}\lambda_{i}}+A_{s}h_{s}S H_{1} H_{2}
+A_{Q}h_{Q}u^{c}_{3} Q_{3} H_{2}+A_{D}h_{D} S D \bar{D} +H.C.)
+m_{3}^{2}|H_{3}|^{2} \nonumber \\
 &  &	+
\sum_{i}{m_{1i}^{2}|H_{1i}|^{2}}+\sum_{i}{m_{2i}^{2}|H_{2i}|^{2}}+\sum_{i}{m_{Si}^{2}|S_{i}|^{2}}+\sum_{i}{m_{Qi}^{2}|Q_{i}|^{2}}+\sum_{i}{m_{ui}^{2}|u_{i}|^{2}}
\nonumber \\
 &  &
+\sum_{i}{m_{di}^{2}|d_{i}|^{2}}+\sum_{i}{m_{Li}^{2}|L_{i}|^{2}}+\sum_{i}{m_{Ei}^{2}|E_{i}|^{2}}+\sum_{i}{m_{Di}^{2}|D_{i}|^{2}}+\sum_{i}{m_{\bar{D}i}^{2}|\bar{D}_{i}|^{2}}
\, . 
\end{eqnarray}

where $i$ is the family index, $\lambda_{i}$ are gauginos, $A_{s}h_{s}$, $A_{Q}h_{Q}$ and $A_{D}h_{D}$ are coefficients of trilinear scalar couplings, and the rest are mass terms for the scalar components of the chiral supermultiplets. 

The tree level scalar potential for $H_{1}$, $H_{2}$ and $S$ has contributions from $F$-terms, $D$-terms and $L_{SB}$:  
\begin{equation}
V=V_{F}+V_{D}+V_{soft}, 
\end{equation}
where, 
\begin{equation}
V_{F}=|h_{s}|^{2}[|H_{1} H_{2}|^{2}+|S|^2(|H_{1}|^2+|H_{2}|^2)]; 
\end{equation}
\begin{equation}
V_{D}=\frac{g_{2}^{2}}{2}|H_{1}^{+}H_{2}|^{2}+\frac{G^{2}}{8}\left( |H_{2}|^{2}-|H_{1}|^{2} \right) ^{2}+\frac{g_{1}^{'2}}{2}\left( Q_{1}|H_{1}|^{2}+Q_{2}|H_{2}|^{2}+Q_{S}|S|^{2} \right) ^{2};
\end{equation}
\begin{equation}
V_{soft}=m_{1}^{2}|H_{1}|^{2}+m_{2}^{2}|H_{2}|^{2}+m_{S}^{2}|S|^{2}-(A_{s}h_{s}SH_{1}H_{2}+H.C.)
\end{equation}
where $G^{2}=g_{2}^{2}+\frac{3}{5}g_{1}^{2}$. The vacuum expectation values of
the Higgs fields and the singlet at the minimum of the scalar potential are
$\langle H_{2}^{+} \rangle=\langle H_{1}^{-}\rangle=0$, $\langle H_{2}^{0}
\rangle=v_{2}/\sqrt{2}$, $\langle H_{1}^{0} \rangle=v_{1}/\sqrt{2}$ and $\langle S \rangle=s/\sqrt{2}$, where $v_{1}$ and $v_{2}$ can be chosen to be real and positive; $s$ is real, and $A_{s}h_{s}s>0$ at the true minimum. 

The minimization conditions for the scalar potential with non-zero VEVs are given by  
\begin{equation}
2m_{S}^{2}-\frac{\sqrt{2}A_{s}h_{s}v_{1}v_{2}}{s}+|h_{s}|^{2}(v_{1}^{2}+v_{2}^{2})+g_{1^{'}}^{2}Q_{S}(Q_{1}v_{1}^{2}+Q_{2}v_{2}^{2}+Q_{S}s^{2})=0;
\end{equation}
\begin{equation}
2m_{2}^{2}-\frac{\sqrt{2}A_{s}h_{s}sv_{1}}{v_{2}}+|h_{s}|^{2}(v_{1}^{2}+s^{2})+g_{1^{'}}^{2}Q_{2}(Q_{1}v_{1}^{2}+Q_{2}v_{2}^{2}+Q_{S}s^{2})=0;
\end{equation}
\begin{equation}
2m_{1}^{2}-\frac{\sqrt{2}A_{s}h_{s}s v_{2}}{v_{1}}+|h_{s}|^{2}(s^{2}+v_{2}^{2})+g_{1^{'}}^{2}Q_{1}(Q_{1}v_{1}^{2}+Q_{2}v_{2}^{2}+Q_{S}s^{2})=0;
\end{equation}
$v^{2}=v_{1}^{2}+v_{2}^{2}=(246~GeV)^{2}$ to ensure the correct electroweak breaking scale. $\tan{\beta}$ is defined as $v_{2}/v_{1}$. The effective $\mu$-term is $\mu_{s}=h_{s}s/\sqrt{2}$.       

After the electroweak and $U(1)^{'}$ symmetries are broken, there are two neutral and massive gauge bosons $Z$ and $Z^{'}$, the $Z-Z^{'}$ mass-squared matrix is given by 
\begin{equation}
(M^{2})_{Z-Z^{'}}= \left( \begin{array}{cc}
M_{Z}^{2} & \Delta ^{2} \\
\Delta^{2} & M_{Z^{'}}^{2} 
\end{array}\right), 
\end{equation}
in which,
\begin{equation}
M_{Z}^{2}=\frac{1}{4}G^{2}(v_{1}^{2}+v_{2}^{2}); 
\end{equation}
\begin{equation}
M_{Z^{'}}^{2}=g_{1^{'}}^{2}(Q_{1}^{2}v_{1}^{2}+Q_{2}^{2}v_{2}^{2}+Q_{S}^{2}s^{2});
\end{equation}
\begin{equation}
\Delta^{2}=\frac{1}{2}g_{1^{'}}G(Q_{1}v_{1}^{2}-Q_{2}v_{2}^{2}). 
\end{equation}
The mass eigenstates are $Z_{1}$ and $Z_{2}$ with 
\begin{equation}
M^{2}_{Z_{1}, Z_{2}}= \frac{1}{2}\left[ M_{Z}^{2}+M_{Z^{'}}^{2}\mp \sqrt{(M_{Z}^{2}-M_{Z^{'}}^{2})^{2}+4\Delta ^{4}} \right]. 
\end{equation}
The $Z-Z^{'}$ mixing angle is given by 
\begin{equation}
\alpha_{Z-Z^{'}}=\frac{1}{2}\arctan{(\frac{2\Delta^{2}}{M_{Z}^{2}-M_{Z^{'}}^{2}})}. 
\end{equation}

Present direct searches and precision tests suggest that $M_{Z}=Gv/2=91.2~GeV$, the mixing angle $\alpha$ is less than a few times $10^{-3}$ and $M_{Z^{'}}>500~GeV$. It has been argued that for certain (e.g., leptophobic) models, a lighter $Z^{'}$ is allowed.\footnote{These include leptophobic models in which $Z^{'}$ couplings to ordinary leptons are absent \cite{s8}; fermiophobic models with no couplings to ordinary fermions \cite{s8f}; and models in which the $Z^{'}$ only couples to the third family \cite{s83}.} In principle, there are also bounds on the parameters (e.g. $A_{i}$, $h_{s}$ etc.) from searches for physical Higgs, exotic particles, gauginos etc. However, it is not our purpose to construct a fully realistic model or consider the detailed mass spectrum for all of the scalar particles in the theory. Hence, we simplify by taking the limits on the masses of scalar particles at low energy to be $>100~GeV$, the gaugino masses to be $>50~GeV$ and $m_{h^{0}}>90~GeV$ as rough phenomenological constraints.

It was pointed out in \cite{s3} that there are two scenarios which can give desired low energy phenomenology: 

$\bullet$ The large $A_{s}$ scenario predicts a small $Z-Z^{'}$ mixing angle and yields a lighter $Z^{'}$ mass. When the $A_{s}$ term is dominant in the scalar potential, $V_{D}$ pushes the minimum to take place at $v_{1}\sim v_{2}$ (second term in $V_{D}$), and $v_{1} \sim v_{2}\sim s$ (third term in $V_{D}$) since $Q_{1}+Q_{2}+Q_{S}=0$. As a result, $\alpha_{Z-Z^{'}}$, proportional to $Q_{1}v_{1}^{2}-Q_{2}v_{2}^{2}$, is small in the case $Q_{1}=Q_{2}$. This scenario is only acceptable if the $Z^{'}$ has suppressed couplings to ordinary particles, as in leptophobic models. 

$\bullet$ The large $\langle S \rangle$ scenario occurs when $s \gg v_{1},~v_{2}$ due to cancellations and therefore $M_{Z^{'}} \gg M_{Z}$, i.e., the $U(1)^{'}$ symmetry breaking occurs at an energy scale higher than the electroweak scale. It was argued in \cite{s3} that $M_{Z^{'}}$ could be of order $TeV$. The first minimization equation gives the singlet VEV,
\begin{equation}
s^{2} = -\frac{2m_{S}^{2}}{g_{1^{'}}^{2}Q_{S}^{2}}+{\cal O}(M_{Z}^{2}). 
\end{equation}
The other two minimization equations can be solved for $v_{1}$ and $v_{2}$,
\begin{equation}
2m_{2}^{2}-\sqrt{2}A_{s}h_{s}s\frac{v_{1}}{v_{2}}+(|h_{s}|^{2}+g_{1^{'}}^{2}Q_{2}Q_{S})s^{2}={\cal O}(M_{Z}^{2}); 
\end{equation}
\begin{equation}
2m_{1}^{2}-\sqrt{2}A_{s}h_{s}s\frac{v_{2}}{v_{1}}+(|h_{s}|^{2}+g_{1^{'}}^{2}Q_{1}Q_{S})s^{2}={\cal O}(M_{Z}^{2}); 
\end{equation}
where the RHS of the equations are functions of $v_{1}$ and $v_{2}$. As $|m_{1}^{2}|$, $|m_{2}^{2}|$ and $|m_{S}^{2}|$ are expected to be the same order of magnitude $\sim {\cal O}(TeV)^{2}$ at $M_{Z}$, cancellations are necessary for the LHS of the equations to have solutions for the Higgs VEVs that yield the correct electroweak scale ($\sqrt{v_{1}^{2}+v_{2}^{2}}=246~GeV$). After symmetry breaking, $M_{Z^{'}} \sim \sqrt{-2m_{S}^{2}}$, and $\alpha_{Z-Z^{'}}\sim G(v_{1}^{2}Q_{1}-v_{2}^{2}Q_{2})/(2g_{1^{'}}Q_{S}^{2}s^{2}) \ll 1$.  \\    

\section{RGE analysis}

Before discussing the symmetry breakings, we briefly describe the renormalization group equation analysis of the model. (The complete set of (1-loop) RGEs is presented in Appendix A.)

(i) Gauge unification. As we have argued in the last section, the supersymmetrized $3 \times 27_{L}+(27_{L}+27_{L}^{*})|_{h_{2}+h_{3}}$ of $E_{6}$ is consistent with gauge unification. The unification scale $M_{G}$ is that at which the gauge coupling constants $g_{2}(\mu)$ for $SU(2)_{L}$ and $g_{1}(\mu)$ for $U(1)_{Y}$ meet, starting with their experimental values at $M_{Z}$. For this model, and working at the 1-loop level, which is sufficient for our purpose, $M_{G}\sim 2\times 10^{16}~GeV$, and $g_{3}(M_{G})=g_{2}(M_{G})=g_{1}(M_{G})=g_{1^{'}}(M_{G})=g_{0}=1.20$. This differs from the value $0.71$ found in the MSSM due to the exotic $(5,~-2)_{L}+(5^{*},~2)_{L}$ representations, which don't affect the unification itself (at 1-loop) or the value of $M_{G}$, but affect the value of the coupling at $M_{G}$ \cite{s4A}\cite{s11}. As a consistency check, the running of $g_{3}$ from $M_{G}$ down to $M_{Z}$ yields $g_{3}^{2}(M_{Z})/4\pi=0.114$, within $5\%$ of the experimental value \cite{s4A}. (The 1-loop $\beta$ function of $g_{3}$ for this model is zero.) The small inconsistency between $M_{G}$ and the value $\sim 5\times 10^{17}~GeV$ expected in perturbative string models, which is $<10\%$ in the $log$, is not significant for the issues considered in this paper.   

(ii) Yukawa couplings. Inspired by the predictions of free fermionic constructions of string theory, we assume that the Yukawa couplings are also unified at $M_{G}$, and for simplicity, their values are taken to be $h_{s}(M_{G})=h_{Q}(M_{G})=h_{D}(M_{G})=h_{0}=g_{0}$ \cite{s3}.   

(iii) The soft supersymmetry breaking parameters at $M_{G}$ are written in terms of dimensionless parameters $c_{i}$, $c_{Ai}$ and $c_{1/2i}$
\begin{equation}
m_{i}^{2}=c_{i}^{2}M_{0}^{2};~ A_{i}=c_{Ai}M_{0};~ M_{i}=c_{1/2i}M_{0},
\end{equation}
where $M_{0}={\cal O}(TeV)$. In each case, $M_{0}$ is rescaled after running everything to low energy to yield $v=246~GeV$. 

(iv) Kinetic mixing. It was shown \cite{s8B} \cite{s8C} that the gauge fields of two $U(1)$'s can be mixed through a term in the Lagrangian which is consistent with all the symmetries in the theory. In our model, the pure kinetic energy terms of the $U(1)_{Y}$ and $U(1)^{'}$ gauge fields can be written as 
\begin{equation}
-L=\frac{1}{4}F_{Y}^{\mu \nu}F_{Y\mu \nu}+\frac{1}{4}F^{'\mu \nu}F^{'}_{\mu \nu}+\frac{\sin{\xi}}{2}F_{Y}^{\mu \nu}F^{'}_{\mu \nu}, 
\end{equation}
where $F_{Y}^{\mu \nu}$ and $F^{'\mu \nu}$ are the field strength for $U(1)_{Y}$ and $U(1)^{'}$, in the basis of the fields in which the interaction terms have the canonical form. If $\xi$ is initially zero, it can arise from loop effects, when $TrQ_{Y}Q \not= 0$, where the trace is restricted to the states lighter than the energy scale being considered. For the $\eta$ model, $TrQ_{Y}Q_{\eta} =0.8$; for the $\psi$ model, it is $-2/\sqrt{10}$. In both cases, the non-zero value is due to the decoupling of everything but two Higgs doublets in the extra $27_{L}+27^{*}_{L}$. Therefore, the kinetic mixing must be included in RGE analysis. 

A non-unitary transformation on the two $U(1)$ gauge fields can be introduced \cite{s8A},
\begin{equation}
A_{Y \mu}=A_{1\mu}-\tan{\xi}A_{2\mu};~~ 
A_{\mu}^{'}=A_{2\mu}/\cos{\xi},
\end{equation}
to diagonalize the kinetic energy terms. In the new basis of the gauge fields, the interaction terms of the chiral fields are 
\begin{equation}
L_{int}=\bar{\psi_{i}}\gamma^{\mu}[g_{1}Q_{Yi}A_{1\mu}+(g_{1^{'}}Q^{'}_{i}+g_{12}Q_{Yi})A_{2\mu}]\psi_{i}, 
\end{equation}
where the redefined gauge coupling constants, written in terms of the original ones, are
\begin{equation}
g_{1}=g_{1}^{0}; ~ g_{1^{'}}=g_{1^{'}}^{0}/\cos{\xi}; ~ g_{12}=-g_{1}^{0}\tan{\xi}. 
\end{equation}
Effectively, the kinetic mixing changes the $U(1)^{'}$ charges of the fields to $Q_{eff}=Q+\delta Q_{y}$, where $\delta \equiv g_{12}/g_{1^{'}}$, while the $U(1)_{Y}$ charges remain the same. The renormalization group equations for the couplings $g_{1}$, $g_{1^{'}}$ and $g_{12}$ are written in Appendix A. As the gauge coupling constants are scale dependent, the effective $U(1)^{'}$ charges are scale dependent as well.  

\section{universal boundary conditions} 

We first consider universal soft supersymmetry breaking.
The universal boundary conditions are: 

${\bullet}$universal scalar masses, 
\begin{equation}
m_{i}^{2}=M_{0}^{2}; 
\end{equation}
${\bullet}$universal gaugino masses, 
\begin{equation}
M_{1}=M_{2}=M_{3}=M_{1}^{'}=c_{1/2}M_{0}; 
\end{equation}
${\bullet}$universal trilinear couplings,
\begin{equation}
A_{s}=A_{Q}=A_{D}=c_{A}M_{0}. 
\end{equation}
The RGEs are solved numerically for the running of all of the soft parameters. The running of the Yukawa couplings, trilinear couplings, and the soft mass square parameters are shown in Figure 1 and 2, with $c_{A}=c_{1/2}=1$. These graphs illustrate the general features that $h_{s}$ is driven down much faster than $h_{Q}$ and $h_{D}$ (because $h_{s}$ is the coupling constant of $SH_{1}H_{2}$, it receives contributions from both $h_{Q}$ (through $H_{2}$) and $h_{D}$ (through $S$)). Thus, $h_{s}(M_{Z})\sim 0.22$, while $h_{Q}(M_{Z})$ $\sim$ $h_{D}(M_{Z})$ $\sim 1.3$. Similarly, $A_{s}(M_{Z})$ is negative and much larger in magnitude than $ A_{Q},~A_{D}$. $m_{2}^{2}$ and $m_{S}^{2}$ are both driven to be negative at low energy, due to the Yukawa couplings to the top quark and the exotic quark, while all other soft mass squares remain positive at $M_{Z}$. Gaugino masses directly affect the running of the trilinear couplings and the soft masses. In most cases, the competition between $A_{i}$'s and $M_{i}$'s controls the differences between the soft mass squared parameters at low energy. Only the scalar fields $S$, $H_{1}$ and $H_{2}$ associated with the non-zero terms in the superpotential (2.6) play a role in low energy symmetry breaking. 

In Figure 3, $M_{Z^{'}}$ at low energy is plotted as a function of $c_{A}$
($c_{A}$ varies over a wide range for different value of $c_{1/2}$, with the
range chosen so that the color symmetry is not broken.). Figure 4 is a plot of
the $Z-Z^{'}$ mixing angle $\alpha$ vs. $c_{A}$ for different
$c_{1/2}$. Figure 3 shows that with universal boundary conditions, $M_{Z^{'}}
< M_{Z}$, e.g., $M_{Z^{'}} \sim 85~GeV$ for the $\psi$ model and
$M_{Z^{'}}\sim 65~GeV$ for the $\eta$ model. Figure 4 shows that
$\alpha_{Z-Z^{'}}$ is not small in general, e.g., $|\alpha_{Z-Z^{'}}| > 0.4$ for the $\eta$ model. The $\psi$ model yields a smaller $Z-Z^{'}$ mixing angle, but still larger than is allowed experimentally, unless $c_{A}$ is large. The scalar potential tends to push $v_{1}$ to be close to $v_{2}$ at the minimum, while the $Z-Z^{'}$ mixing angle $\alpha \propto (Q_{1}v_{1}^{2}-Q_{2}v_{2}^{2})$. Compared with the $\psi$ model, which has $Q_{1}=Q_{2}$, the $\eta$ model with $Q_{2}=4Q_{1}$ naturally yields a larger mixing angle. 

The plot shows that small mixing angles ($\sim 0.01$) are possible for the $\psi$ model when $c_{A}$ is large ($>10$) and the gaugino masses are small. This case is basically the large $A_{s}$ scenario described in \cite{s3}, in which the $A_{s}$ term ($A_{s}h_{s}SH_{1}H_{2}$) dominates the scalar potential and pushes the minimum to $v_{1}=v_{2}=s$. With $Q_{1}=Q_{2}$ for the $\psi$ model, the limiting case has a vanishing $Z-Z^{'}$ mixing angle. However, the $Z^{'}$ mass, controlled by $
g_{1}^{'2}(Q_{1}^{2}v_{1}^{2}+Q_{2}^{2}v_{2}^{2}+Q_{s}^{2}s^{2})$, is not acceptable, because $g_{1^{'}}$ is smaller than $g_{2}$ at electroweak scale and the $U(1)^{'}$ charges are smaller than the $U(1)_{Y}$ charges. As a result, $M_{Z^{'}}<M_{Z}$, which is excluded experimentally..

The large $\langle S \rangle$ scenario cannot be realized with universal
boundary conditions in either of the $E_{6}$ models\footnote{If Yukawa
universality is broken, especially if $h_{s}$ is much larger than
$h_{U}$ and $h_{D}$ at the string scale, $h_{s}(M_{Z})$ can be large enough to
make $(|h_{s}|^{2}+g_{1^{'}}^{2}Q_{2}Q_{S})$ positive in (2.22), so that the
cancellation equations for the large $\langle S \rangle$ scenario could be
satisfied with universal soft supersymmetry breaking parameters. Discussions
in \cite{s13} belongs to this category. In string models based on the fermionic
($Z_{2}\times Z_{2}$) orbifold construction \cite{s5A}\cite{s5B}\cite{s5C},
the couplings of the trilinear terms in the superpotential can be
$g_{0}$, $g_{0}/\sqrt{2}$ or $g_{0}/2$, depending on the
number of Ising fermion excitations involved in the string vertex operators
\cite{s3}. In this case, the maximum possible splitting between $h_{s}$ and
$h_{U}$, $h_{D}$ at the string scale is indeed large enough to make
$(|h_{s}|^{2}+g_{1^{'}}^{2}Q_{2}Q_{S})$ positive, for both the $\eta$ and
$\psi$ models, to induce the large $s$ solution at low energy. Even larger
splittings are possible for effective Yukawa couplings derived from higher
dimensional operators \cite{s13A}.} In the $\psi$ model, $Q_{1}=Q_{2}=-2/\sqrt{24}$, $Q_{S}=4/\sqrt{24}$, $h_{s}(M_{Z})\sim 0.2$, and $g_{1^{'}}(M_{Z})\sim 0.45$ so that $(|h_{s}|^{2}+g_{1}^{'2}Q_{2}Q_{S})<0$. Since the $A_{s}h_{s}s$ term is always positive at the true minimum, the cancellation conditions (2.22) and (2.23) cannot be satisfied with negative $m_{2}^{2}$. For the same reason, the $\eta$ model fails to yield large $\langle S \rangle$ solutions with universal boundary conditions.

Finally, we comment on the effect of kinetic mixing. It was argued \cite{s8A} that for the $\eta$ model, with an additional pair of Higgs doublets chosen from the $78$ representation of the $E_{6}$ group, $\delta(M_{Z})$ can be large enough to make the $Z^{'}$ approximately leptophobic. Thus, a lighter $Z^{'}$ could possibly be allowed at low energy. In our model, the additional Higgs doublets come from a $27+27^{*}$. We find that 
$\delta(M_{Z})\sim 0.08$ for the $E_{6}$ $\eta$ model, while $\delta(M_{Z})$ needs to be $1/3$ to make the $\eta$ model leptophobic. Therefore, the kinetic mixing is only a small effect, and $Z^{'}$ is not leptophobic. Similarly, kinetic mixing is too small to reverse the sign of the $U(1)^{'}$ charges so that the cancellation conditions (2.22) and (2.23) can be satisfied with negative $m_{2}^{2}$.\\

\section{nonuniversal boundary conditions}

To have acceptable low energy phenomenology for both the $\eta$ and the $\psi$ model, we invoke nonuniversal boundary conditions
to satisfy the conditions (2.22) and (2.23) for large $\langle S \rangle$
solutions. We keep universal gaugino masses and universal trilinear couplings:
$M_{i}=c_{1/2}M_{0}$, $A_{i}=c_{A}M_{0}$, as well as Yukawa universality. Among the soft mass squared parameters, we adjust only $m_{1}^{2}$, $m_{2}^{2}$, $m_{S}^{2}$, $m_{Q_{3}}^{2}$, $m_{u^{c}_{3}}^{2}$, $m_{D}^{2}$ and $m_{\bar{D}}^{2}$, i.e., those that dominate the running of the RGEs.  

The purpose is to adjust the soft mass squared parameters at $M_{G}$ so that
$m_{2}^{2}$ and $m_{1}^{2}$ are both positive at low energy, while $m_{S}^{2}$
is negative. $|m_{1}^{2}|$, $|m_{2}^{2}|$ and $|m_{S}^{2}|$ should be fairly
close to each other to satisfy the cancellation equations, so we must choose
the gaugino masses to be small, while ensuring that the low energy values of
the chargino and gluino masses satisfy existing experimental bounds. The
trilinear coupling $A_{s}$ must also be small at $M_{Z}$ to avoid the other
extreme, the large $A_{s}$ scenario. We adjust the squark masses at $M_{G}$ to
ensure that the color symmetry is not broken at the electroweak scale; i.e,
all squark masses must be positive (including exotics), and $A_{Q}$, $A_{D}$
must not be too large \footnote{A large $A_{Q}$, which implies that the global
minimum of the potential breaks charge and color, may be acceptable if the
charge/color conserving minimum is populated first cosmologically and is sufficiently long-lived \cite{s14}}. The sparticle masses at $M_{Z}$ also need to satisfy the phenomenological bounds. 

In Tables 3 and 4, we give a few examples of nonuniversal boundary conditions
that induce the large $\langle S \rangle$ scenario at $M_{Z}$, for both the
$\psi$ and the $\eta$ model. For each example, the first row for each soft
supersymmetry breaking parameter gives its value at the high scale $M_{G}$,
while the second row gives its value at $M_{Z}$. Only the gaugino masses $M_{2}$ and $M_{1}^{'}$ are listed, because $M_{3}$ is a constant due to the vanishing 1-loop $\beta$ function for $SU(3)_{c}$, and $M_{1}$ is always only slightly larger than $M_{1}^{'}$. $A_{s}$ and $A_{Q}$ are given, and $A_{D} \sim A_{Q}$. The $Z^{'}$ mass, $Z-Z^{'}$ mixing angle, and $\tan{\beta}$ are given for each case.

Some general patterns can be seen from these examples. For example,
$m_{2}^{2}$ has to be larger than the other mass squared parameters at $M_{G}$
so that it can be positive at the electroweak scale. $m_{S}^{2}$ has to be
increased to decrease the difference between $|m_{2}^{2}|$ and
$|m_{S}^{2}|$. Squark masses have to be adjusted accordingly to avoid color
symmetry breaking at $M_{Z}$. Generally, the trilinear couplings are large and negative, so that $A_{s}$ can be small at low energy.  

Compared with the $\psi$ model, the $\eta$ model yields solutions with large $\tan{\beta}$ ($>1$) and a slightly larger $Z-Z^{'}$ mixing angle. The $U(1)^{'}$ charge assignments of the particles in the theory affect the low energy phenomenology in a nontrivial way.\\

\section{mixing between the two $U(1)^{'}$s in $E_{6}$}

There is another way to satisfy the cancellation equations (2.22) and (2.23)
for the large $\langle S \rangle$ minimum. As we have seen, negative
$m_{2}^{2}$ and $m_{S}^{2}$ with positive $m_{1}^{2}$ at low energy are generic to our models with quarks that couple to $H_{2}$ and exotics that couple to $S$ through large Yukawa couplings with universal boundary conditions, while $m_{1}^{2}$ remains positive. Yet, eqn. (2.22) and (2.23) could be satisfied if $Q_{1}Q_{S}<0$ and $Q_{2}Q_{S}>0$. These conditions do not hold for the $\psi$ and $\eta$ models even with kinetic mixing, illustrating the strong $U(1)^{'}$ charge influence on the final $Z^{'}$ mass and $\alpha_{Z-Z^{'}}$. The $U(1)^{'}$ charges affect the running of the RGEs only slightly; $\alpha_{Z-Z^{'}}$ is affected through the $U(1)^{'}$ charge dependence of the $Z-Z^{'}$ mass-squared matrix. However, the largest effect of the $U(1)^{'}$ charges comes from the minimization of the scalar potential.

In a general $E_{6}$ model with a single extra $U(1)^{'}$, the $U(1)^{'}$ charge can be combination of $U(1)_{\psi}$ and $U(1)_{\chi}$ with 
\begin{equation}
Q=Q_{\chi}\cos{\theta_{E_{6}}}+Q_{\psi}\sin{\theta_{E_{6}}}, 
\end{equation}   
where $\theta_{E_{6}}$ is a mixing angle which can be chosen to be in the
range $0$-$\pi$. If $\cos{\theta_{E_{6}}}>\sqrt{3/5}\sim 0.79$,  
one has $Q_{1}Q_{S}<0$ and $Q_{2}Q_{S}>0$.

With universal soft supersymmetry breaking parameters at the large scale, this model has four free parameters: $\theta_{E_{6}}$, $M_{0}$, $c_{A}$ and $c_{1/2}$. $s$ is fixed by (2.21), and the other two equations, taken as cancellation conditions for a large $s$ minimum, are to be satisfied. Our strategy is to take a particular gaugino mass, vary both $c_{A}$ and $\theta_{E_{6}}$, search for large $s \gg v_{1}, ~v_{2}$ at low energy, then rescale $M_{0}$.   

As an example, we show the parameter ranges of $(c_{A},~\theta_{E_{6}})$ for
$M_{Z^{'}} > 500~GeV$ as a 2-D contour plot in Figure 5, taking $c_{1/2}=0.4$
for the gaugino mass. As we have argued, gaugino masses need to be small to
avoid huge splitting between the mass squares at low energy, and for smaller
$c_{1/2}$, smaller $Q_{i}Q_{S}$ is needed for the cancellation so that large
$s$ occurs with smaller $\cos{\theta_{E_{6}}}$. If $|A_{s}|$ is large
($A_{s}(M_{Z})$ should be small to avoid a large $A_{s}$ minimum), a large
value of $\cos{\theta_{E_{6}}}$ is necessary to increase the $Q_{i}/Q_{S}$ ratio for the cancellation. The curves are terminated at the ends to preserve the color symmetry.

The curves are approximately symmetric around $c_{A}=-4$. In the limit in
which the gaugino masses are neglected in the RGEs, the symmetry can be seen
as the follows: consider the cases 1 and 2 with $c_{A}=\pm c_{0}$ at the large scale. The RGEs of $A_{i}$ tell us that at $A_{s}(t)=0$, 
\begin{equation}
\frac{d}{dt}A_{s}=6A_{Q}h_{Q}^{2}+6A_{D}h_{D}^{2} \sim 12 A_{Q}h_{Q}^{2},
\end{equation}
where we have assumed that $h_{Q}\sim h_{D}$, $A_{Q}\sim A_{D}$. Also, $A_{Q}$
and $A_{D}$ stay close to their initial values at $M_{G}$. Hence, at the point
that $A_{s}$ crosses zero, the slopes of $A_{s}$ for cases 1 and 2 are
approximately equal and opposite, which induces the equal and opposite $A_{s}$ for cases 1 and 2 at $M_{Z}$. The running of the soft mass squared parameters only depend on $A_{i}^{2}$, so that $m_{i}^{2}(1) \sim m_{i}^{2}(2)$ at the electroweak scale. Hence, equal and opposite initial conditions of $A_{i}$ result in the same low energy phenomenology. However, the contribution of the gaugino masses shift the center of the symmetry from $c_{A}=0$.              

\section{summary}

In this paper, we presented a detailed analysis of the electroweak and
$U(1)^{'}$ symmetry breaking for a supersymmetric model with the matter
content $3\times 27_{L}+(27_{L}+27_{L}^{*})|_{h_{2}+h_{3}}$ of $E_{6}$. It has
an anomaly free $U(1)^{'}$ symmetry, is consistent with gauge unification, and
is meant to be a concrete example of the more general scenarios discussed in
\cite{s3}. It includes the three family standard model particle structure and
a pair of Higgs doublets; in addition, there are three exotic quark singlet
pairs $D$ and $\bar{D}$, two extra Higgs doublets $H_{1}$, three extra Higgs
doublets $H_{2}$, and one Higgs doublet $H_{3}$, which is the conjugate of
$H_{2}$. The extra $U(1)^{'}$ symmetry is taken to be $U(1)_{\psi}$ or
$U(1)_{\eta}$ of the $E_{6}$ model, or a combination of $U(1)_{\psi}$ and
$U(1)_{\chi}$. Couplings between the singlet $S$ and Higgs doublets $H_{1}$,
$H_{2}$ and couplings between the singlet and the exotic quarks $D$, $\bar{D}$
are naturally allowed by gauge invariance. The model is string (rather than
GUT) motivated. Thus, some terms in the superpotential that would be allowed
by gauge invariance are assumed to be absent due to string selection rules. In
particular, we assume the absence of baryon and lepton number violating
couplings of the exotic $D$, $\bar{D}$ quarks, so that $D$ and $\bar{D}$ can
be light. Similarly, we assume that most of the possible Yukawa couplings of
the Higgs doublets and singlets vanish, so that only $m_{t}$ is large and two
higgs doublets and one singlet participate in the symmetry breaking. We make
the string motivated assumption that the non-zero Yukawa couplings are equal
at the string scale (Yukawa universality), and use the value $\sim
g_{0}$ for definiteness.

With the specific particle content and $U(1)^{'}$ charge assignments, we use the numerical results of the RGE analysis to explore features of the radiative symmetry breaking at the electroweak scale. Our results are summarized as the follows. 

$\bullet$ Only the Higgs doublets and singlet with non-zero Yukawa couplings play a role in symmetry breaking.

$\bullet$ The symmetry breaking scenarios yield an effective $\mu$ term at low energy, $\mu_{eff}=h_{s}\langle S \rangle$.

$\bullet$ The universal soft supersymmetry breaking mass parameters at the large scale fail to give phenomenologically acceptable scenarios at low energy for the $E_{6}$ $\psi$ and $\eta$ models.  

$\bullet$ To achieve the desired low energy phenomenology for the
supersymmetric $E_{6}$ model, one has to look for solutions with large $S$
VEVs, i.e., $\langle S \rangle \gg \langle H_{1} \rangle , \langle H_{2}
\rangle$. There are two ways to reach such solutions while maintaining Yukawa universality.  

(i) For both the $E_{6}$ $\psi$ and $\eta$ models, by changing the soft supersymmetry breaking mass squared parameters at the unification scale, there are parameter regions that yield large $s$ at the electroweak scale, with a heavy $Z^{'}$ and small $Z-Z^{'}$ mixing angle.

(ii) By introducing a mixing angle $\theta_{E_{6}}$ between the two extra $U(1)$ symmetries of the $E_{6}$ model, one can vary the $U(1)^{'}$ charge assignments of the particles in the model, while keeping the model anomaly free. With universal boundary conditions at the unification scale, for given gaugino masses, there exist parameter ranges of $c_{A}$ (trilinear couplings) and $\theta_{E_{6}}$ at $M_{G}$ that yield the large $s$ minimum at low energy. 

We confirm the two scenarios that may induce acceptable low energy phenomenology, the large $A_{s}$ scenario and large $\langle S \rangle$ scenario, proposed in Ref. \cite{s3}. The large $A_{s}$ scenario fails in this case to give phenomenologically acceptable $Z^{'}$ masses in the $E_{6}$ $\psi$ and $\eta$ models because the couplings are not leptophobic, even including kinetic mixing. It is nevertheless a useful example of this scenario, which maybe viable in other string derived models with suppressed couplings to ordinary fermions. The large $\langle S \rangle$ scenario can be realized, either through non universal boundary conditions for either the $\psi$ or $\eta$ model or by varying the $U(1)^{'}$ quantum numbers of the particles.

\section{acknowledgments}

This work was supported in part by U.S. Department of Energy Grant No. DOE-EY-76-02-3071. It is a pleasure to thank M. Cveti\v c, G. Cleaver, J. R. Espinosa, L. Everett and J. Erler for useful discussions.  

\section{Appendix A: renormalization group equations}

We present here the complete one loop renormalization group equations of the anomaly-free $E_{6}$ model: $3\times 27_{L}+(27_{L}+27_{L}^{*})|_{h_{2}+h_{3}}$, based on the particle content as listed in Table 2. In these equations, the scale variable is defined as
\begin{equation}
t=\frac{1}{16\pi^{2}}\ln \frac{\mu}{M_{G}},
\end{equation}
where $M_{G}$ is determined to be $2 \times 10^{16}~GeV$ by the running of the $SU(2)_{L}$ gauge coupling constant $g_{2}(\mu)$ and the $U(1)_{Y}$ gauge coupling constant $g_{1}(\mu)$, with their inputs from $\mu=M_{Z}$.  \\

{\bf Gauge couplings:} 
\begin{equation}
\frac{d}{dt}g_{i}=\beta_{i} g_{i}^{3}, 
\end{equation}
where $i=1,~2,~3$ for $U(1)_{Y}$, $SU(2)_{L}$ and $SU(3)_{c}$ respectively. The $\beta$ functions are
\begin{equation}
\beta_{1}= TrQ_{Y}^{2}=\frac{48}{5};~~ \beta_{2}= 4;~~ \beta_{3}= 0. 
\end{equation}
With kinetic mixing, the equations for the $U(1)^{'}$ gauge coupling constant $g_{1^{'}}$ and the $U(1)_{Y}$-$U(1)^{'}$-mixing constant $g_{11^{'}}$ \cite{s8A}(see section III for explanations) are 
\begin{equation}
\frac{d}{dt}g_{1^{'}}=(\beta_{1^{'}}g_{1^{'}}^{2}+\beta_{1}g_{11^{'}}+2\beta_{11^{'}}g_{1^{'}}g_{11^{'}})g_{1^{'}},
\end{equation}
\begin{equation}
\frac{d}{dt}g_{11^{'}}=(\beta_{1^{'}}g_{1^{'}}^{2}+\beta_{1}g_{11^{'}}+2\beta_{11^{'}}g_{1^{'}}g_{11^{'}})g_{11^{'}}+2g_{1}^{2}(g_{11^{'}}\beta_{1}+g_{1^{'}}\beta_{11^{'}}),
\end{equation}   
where, 
\begin{equation}
\beta_{1^{'}}=TrQ^{2};~~ \beta_{11^{'}}=TrQQ_{y};
\end{equation}
The $U(1)^{'}$ charges of the particles $Q$ are $Q_{\psi}$ in the $\psi$ model, $Q_{\eta}$ in the $\eta$ model and $(Q_{\chi}\cos{\theta_{E_{6}}}+Q_{\psi}\sin{\theta_{E_{6}}})$ in the model with mixing between $U(1)_{\chi}$ and $U(1)_{\psi}$. With the kinetic mixing, the effective $U(1)^{'}$ charge is $Q_{eff}=Q+\delta Q_{y}$, with $\delta=g_{11^{'}}/g_{1^{'}}$, so that the $U(1)^{'}$ charges are scale dependent. \\

{\bf Gaugino masses:}
\begin{equation}
\frac{d}{dt}M_{i}=2\beta_{i}g_{i}^{2}M_{i}. 
\end{equation}
where $i=1,~2,~3,1^{'}$ for $U(1)_{Y}$, $SU(2)_{L}$, $SU(3)_{c}$ and $U(1)^{'}$, respectively. Based on the observation that the kinetic mixing effect is small (less than $8 \%$) in our model, we neglect the contributions from kinetic mixing in the runnings of soft supersymmetry breaking parameters.   
\\

{\bf Yukawa couplings:}
\begin{equation}
\frac{d}{dt}h_{Q}=h_{Q}\left[ 6h_{Q}^{2}+h_{s}^{2}-\left( \frac{16}{3}g_{3}^{2}+3g_{2}^{2}+\frac{13}{15}g_{1}^{2}+2g_{1^{'}}^{2}(Q_{u}^{2}+Q_{Q}^{2}+Q_{2}^{2})\right)\right];
\end{equation}
\begin{equation}
\frac{d}{dt}h_{s}=h_{s}\left[ 3h_{Q}^{2}+4h_{s}^{2}+3h_{D}^{2}-\left(3g_{2}^{2}+\frac{3}{5}g_{1}^{2}+2g_{1^{'}}^{2}(Q_{S}^{2}+Q_{1}^{2}+Q_{2}^{2})\right)\right];
\end{equation}
\begin{equation}
\frac{d}{dt}h_{D}=h_{D}\left[ 5h_{D}^{2}+2h_{s}^{2}-\left( \frac{16}{3}g_{3}^{2}+\frac{4}{15}g_{1}^{2}+2g_{1^{'}}^{2}(Q_{D}^{2}+Q_{\bar{D}}^{2}+Q_{S}^{2})\right)\right].
\end{equation}
\\
{\bf Trilinear couplings:}
\begin{equation}
\frac{d}{dt}A_{Q}=12A_{Q}h_{Q}^{2}+2A_{s}h_{s}^{2}-2\left[ \frac{16}{3}g_{3}^{2}M_{3}+3g_{2}^{2}M_{2}+\frac{13}{15}g_{1}^{2}M_{1}+2g_{1^{'}}^{2}(Q_{u}^{2}+Q_{Q}^{2}+Q_{2}^{2})M_{1^{'}}\right];
\end{equation}
\begin{equation}
\frac{d}{dt}A_{s}=6A_{Q}h_{Q}^{2}+8A_{s}h_{s}^{2}+6A_{D}h_{D}^{2}-2\left[3g_{2}^{2}M_{2}+\frac{3}{5}g_{1}^{2}M_{1}+2g_{1^{'}}^{2}(Q_{S}^{2}+Q_{1}^{2}+Q_{2}^{2})M_{1^{'}}\right];
\end{equation}
\begin{equation}
\frac{d}{dt}A_{D}=10A_{D}h_{Q}^{2}+4A_{s}h_{s}^{2}-2\left[ \frac{16}{3}g_{3}^{2}M_{3}+\frac{4}{15}g_{1}^{2}M_{1}+2g_{1^{'}}^{2}(Q_{D}^{2}+Q_{\bar{D}}^{2}+Q_{S}^{2})M_{1^{'}}\right].
\end{equation}
\\
{\bf Soft scalar mass-squared parameters:} 
\begin{eqnarray}
\frac{d}{dt}m_{2}^{2}&=&6(m_{2}^{2}+m_{Q_{3}}^{2}+m_{u_{3}}^{2}+A_{Q}^{2})h_{Q}^{2}+2(m_{2}^{2}+m_{1}^{2}+m_{S}^{2}+A_{s}^{2})h_{s}^{2}
\nonumber \\
 &  & -8\left(
\frac{3}{4}g_{2}^{2}M_{2}^{2}+\frac{3}{20}g_{1}^{2}M_{1}^{2}+Q_{2}^{2}g_{1^{'}}^{2}M_{1^{'}}^{2}\right)+\frac{3}{5}g_{1}^{2}S_{1}+2Q_{2}g_{1^{'}}^{2}S_{1^{'}} 
\, ;
\end{eqnarray}

\begin{eqnarray}
\frac{d}{dt}m_{1}^{2} &=
&2(m_{2}^{2}+m_{1}^{2}+m_{S}^{2}+A_{s}^{2})h_{s}^{2}-8\left(
\frac{3}{4}g_{2}^{2}M_{2}^{2}+\frac{3}{20}g_{1}^{2}M_{1}^{2}+Q_{1}^{2}g_{1^{'}}^{2}M_{1^{'}}^{2}\right)
\nonumber \\
 &  & -\frac{3}{5}g_{1}^{2}S_{1}+2Q_{1}g_{1^{'}}^{2}S_{1^{'}} \, ;
\end{eqnarray}

\begin{equation}
\frac{d}{dt}m_{S}^{2} =6(m_{D}^{2}+m_{\bar{D}}^{2}+m_{S}^{2}+A_{D}^{2})h_{D}^{2}+4(m_{2}^{2}+m_{1}^{2}+m_{S}^{2}+A_{s}^{2})h_{s}^{2}-8Q_{S}^{2}g_{1^{'}}^{2}M_{1^{'}}^{2}+2Q_{S}g_{1^{'}}^{2}S_{1^{'}};
\end{equation}

\begin{eqnarray}
\frac{d}{dt}m_{Q_{3}}^{2}
&=&2(m_{2}^{2}+m_{Q_{3}}^{2}+m_{u_{3}}^{2}+A_{Q}^{2})h_{Q}^{2}-8\left(
\frac{4}{3}g_{3}^{2}M_{3}^{2}+\frac{3}{4}g_{2}^{2}M_{2}^{2}+\frac{1}{60}g_{1}^{2}M_{1}^{2}+Q_{Q}^{2}g_{1^{'}}^{2}M_{1^{'}}^{2}\right)
\nonumber \\
 &  & +\frac{1}{5}g_{1}^{2}S_{1}+2Q_{Q}g_{1^{'}}^{2}S_{1^{'}} \, ;
\end{eqnarray}

\begin{eqnarray}
\frac{d}{dt}m_{u_{3}}^{2} &=&
4(m_{2}^{2}+m_{Q_{3}}^{2}+m_{u_{3}}^{2}+A_{Q}^{2})h_{Q}^{2}-8\left(
\frac{4}{3}g_{3}^{2}M_{3}^{2}+\frac{4}{15}g_{1}^{2}M_{1}^{2}+Q_{u}^{2}g_{1^{'}}^{2}M_{1^{'}}^{2}\right)
\nonumber \\
 &  &-\frac{4}{5}g_{1}^{2}S_{1}+2Q_{u}g_{1^{'}}^{2}S_{1^{'}} \, ;
\end{eqnarray}

\begin{eqnarray}
\frac{d}{dt}m_{D}^{2} &=
&2(m_{D}^{2}+m_{\bar{D}}^{2}+m_{S}^{2}+A_{D}^{2})h_{D}^{2}-8\left(
\frac{4}{3}g_{3}^{2}M_{3}^{2}+\frac{1}{15}g_{1}^{2}M_{1}^{2}+Q_{D}^{2}g_{1^{'}}^{2}M_{1^{'}}^{2}\right) 
\nonumber \\
 &  & -\frac{2}{5}g_{1}^{2}S_{1}+2Q_{D}g_{1^{'}}^{2}S_{1^{'}} \, ;
\end{eqnarray}

\begin{eqnarray}
\frac{d}{dt} m_{\bar{D}}^{2} & = 
&2(m_{D}^{2}+m_{\bar{D}}^{2}+m_{S}^{2}+A_{D}^{2})h_{D}^{2}-8\left(
\frac{4}{3}g_{3}^{2}M_{3}^{2}+\frac{1}{15}g_{1}^{2}M_{1}^{2}+Q_{\bar{D}}^{2}g_{1^{'}}^{2}M_{1^{'}}^{2}\right)
\nonumber \\
 &  & +\frac{2}{5}g_{1}^{2}S_{1}+2Q_{\bar{D}}g_{1^{'}}^{2}S_{1^{'}} \, ;
\end{eqnarray}

\begin{equation}
\frac{d}{dt}m_{L}^{2}= -8\left( \frac{3}{4}g_{2}^{2}M_{2}^{2}+\frac{3}{20}g_{1}^{2}M_{1}^{2}+Q_{L}^{2}g_{1^{'}}^{2}M_{1^{'}}^{2}\right)+\frac{3}{5}g_{1}^{2}S_{1}+2Q_{L}g_{1^{'}}^{2}S_{1^{'}};
\end{equation}

\begin{equation}
\frac{d}{dt}m_{E}^{2}= -8\left(\frac{3}{5}g_{1}^{2}M_{1}^{2}+Q_{E}^{2}g_{1^{'}}^{2}M_{1^{'}}^{2}\right)+\frac{6}{5}g_{1}^{2}S_{1}+2Q_{E}g_{1^{'}}^{2}S_{1^{'}};
\end{equation} 
\begin{equation}
\frac{d}{dt}m_{N}^{2}= -8Q_{N}^{2}g_{1^{'}}^{2}M_{1^{'}}^{2}+2Q_{N}g_{1^{'}}^{2}S_{1^{'}};
\end{equation} 

\begin{equation}
\frac{d}{dt}m_{Q_{1,2}}^{2} =-8\left( \frac{4}{3}g_{3}^{2}M_{3}^{2}+\frac{3}{4}g_{2}^{2}M_{2}^{2}+\frac{1}{60}g_{1}^{2}M_{1}^{2}+Q_{Q}^{2}g_{1^{'}}^{2}M_{1^{'}}^{2}\right) +\frac{1}{5}g_{1}^{2}S_{1}+2Q_{Q}g_{1^{'}}^{2}S_{1^{'}};
\end{equation}

\begin{equation}
\frac{d}{dt}m_{u_{1,2}}^{2} =-8\left( \frac{4}{3}g_{3}^{2}M_{3}^{2}+\frac{4}{15}g_{1}^{2}M_{1}^{2}+Q_{u}^{2}g_{1^{'}}^{2}M_{1^{'}}^{2}\right) -\frac{4}{5}g_{1}^{2}S_{1}+2Q_{u}g_{1^{'}}^{2}S_{1^{'}};
\end{equation}

\begin{equation}
\frac{d}{dt}m_{Hu}^{2}=-8\left( \frac{3}{4}g_{2}^{2}M_{2}^{2}+\frac{3}{20}g_{1}^{2}M_{1}^{2}+Q_{2}^{2}g_{1^{'}}^{2}M_{1^{'}}^{2}\right)+\frac{3}{5}g_{1}^{2}S_{1}+2Q_{2}g_{1^{'}}^{2}S_{1^{'}};
\end{equation}

\begin{equation}
\frac{d}{dt}m_{Hd}^{2}=8\left( \frac{3}{4}g_{2}^{2}M_{2}^{2}+\frac{3}{20}g_{1}^{2}M_{1}^{2}+Q_{1}^{2}g_{1^{'}}^{2}M_{1^{'}}^{2}\right) -\frac{3}{5}g_{1}^{2}S_{1}+2Q_{1}g_{1^{'}}^{2}S_{1^{'}};
\end{equation}

\begin{equation}
\frac{d}{dt}m_{H_{3}}^{2}=-8\left( \frac{3}{4}g_{2}^{2}M_{2}^{2}+\frac{3}{20}g_{1}^{2}M_{1}^{2}+Q_{2}^{2}g_{1^{'}}^{2}M_{1^{'}}^{2}\right)-\frac{3}{5}g_{1}^{2}S_{1}-2Q_{2}g_{1^{'}}^{2}S_{1^{'}};
\end{equation}

\begin{equation}
\frac{d}{dt}m_{S_{1,2}}^{2} =-8Q_{S}^{2}g_{1^{'}}^{2}M_{1^{'}}^{2}+2Q_{S}g_{1^{'}}^{2}S_{1^{'}};
\end{equation}

\begin{equation}
\frac{d}{dt}m_{D_{1,2}}^{2}=-8\left( \frac{4}{3}g_{3}^{2}M_{3}^{2}+\frac{1}{15}g_{1}^{2}M_{1}^{2}+Q_{D}^{2}g_{1^{'}}^{2}M_{1^{'}}^{2}\right)-\frac{2}{5}g_{1}^{2}S_{1}+2Q_{D}g_{1^{'}}^{2}S_{1^{'}};
\end{equation}

\begin{equation}
\frac{d}{dt}m_{\bar{D}_{1,2}}^{2}=-8\left( \frac{4}{3}g_{3}^{2}M_{3}^{2}+\frac{1}{15}g_{1}^{2}M_{1}^{2}+Q_{\bar{D}}^{2}g_{1^{'}}^{2}M_{1^{'}}^{2}\right)+\frac{2}{5}g_{1}^{2}S_{1}+2Q_{\bar{D}}g_{1^{'}}^{2}S_{1^{'}};
\end{equation}
where $m_{Hu}$ and $m_{Hd}$ are the masses of the Higgs doublets that don't have Yukawa couplings. $S_{1}$ and $S_{1^{'}}$ in these equations are defined as

\begin{eqnarray}
S_{1} & = &\sum\limits^{3}_{i=1}
(m_{Q_{i}}^{2}-2m_{u_{i}}^{2}+m_{d_{i}}^{2}-m_{L_{i}}^{2}+m_{E_{i}}^{2}+m_{D_{i}}^{2}+m_{\bar{D}_{i}}^{2})
\nonumber \\
 &  & +m_{2}^{2}-m_{1}^{2}+3m_{Hu}^{2}-2m_{Hd}^{2}-m_{H_{3}}^{2} \, ;
\end{eqnarray}

\begin{eqnarray}
S_{1^{'}} &=&\sum\limits^{3}_{i=1}
(6Q_{Q}^{2}m_{Q_{i}}^{2}+3Q_{u}^{2}m_{u_{i}}^{2}+3Q_{d}^{2}m_{d_{i}}^{2}+2Q_{L}^{2}m_{L_{i}}^{2}+Q_{E}^{2}m_{E_{i}}^{2}+3Q_{D}^{2}m_{D_{i}}^{2}+3Q_{\bar{D}}^{2}m_{\bar{D}_{i}}^{2})
\nonumber \\
 & &
+Q_{S}^{2}m_{S_{i}}^{2}+Q_{N}^{2}m_{N_{i}}^{2})+2Q_{1}(m_{1}^{2}+2m_{Hd}^{2})+2Q_{2}(m_{2}^{2}+3m_{Hu}^{2})-2Q_{2}m_{H_{3}}^{2} 
\, . 
\end{eqnarray}

\bigskip

\bibliographystyle{prsty}
\bibliography{tdw}

\newpage
\begin{table}
\begin{center}
\begin{tabular}{|c|c|c|c|c|}   
$SO(10)$ & $Su(5)$  & $2\sqrt{10}Q_{\chi}$ & $2\sqrt{6}Q_{\psi}$ & $2\sqrt{15}Q_{\eta}$\\ \hline   
$16$ & $10(u, d, \bar{u}, e^{+})_{L}$& $-1$& $1$& $-2$  \\ 
	 & $5^{*}(\bar{d}, \nu, e^{-})_{L}$&$3$& $1$& $1$  \\ 
	    & $1\bar{N}_{L}$&$-5$&$1$&$-5$ \\ \hline
$10$ & $5(D, h_{2}^{0}, h_{2}^{+})_{L}$&$2$& $-2$& $4$  \\
	    & $5^{*}(\bar{D}, h_{1}^{0}, h_{1}^{-})_{L}$&$-2$&$-2$&$1$ \\ \hline
$1$ & $1S_{L}^{0}$& $0$& $4$& $-5$  

\end{tabular}
\caption{Decomposition of $E_{6}$ model $27$-plet representation.}
\end{center}
\end{table}

\begin{table}
\begin{center}
\begin{tabular}{|c|c|c|c|c|}
\small
matter multiplets & $SU(3)_{c}$ & $SU(2)_{L}$ & $U(1)_{Y}$ & $U(1)^{'}$ \\ \hline
$\hat{Q}_{i}$ &  3 & 2 & 1/6 & $Q_{Q}$ \\ \hline 
$\hat{u}^{c}_{i}$ & 3 & 1 & $-2/3$ & $Q_{u}$ \\ \hline
$\hat{d}^{c}_{i}$ & 3 & 1 & 1/3 & $Q_{d}$ \\ \hline 
$\hat{L}_{i}$ & 1 & 2 & $-1/2$ & $Q_{L}$ \\ \hline 
$\hat{E}^{c}_{i}$ & 1 & 1 & 1 & $Q_{E}$ \\ \hline 
$\hat{H}_{2i}$ & 1 & 2 & 1/2 & $Q_{1}$ \\ \hline 
$\hat{H}_{1i}$ & 1 & 2 & $-1/2$ & $Q_{2}$ \\ \hline 
$\hat{D}_{i}$ & 3 & 1 & $-1/3$ & $Q_{D}$ \\ \hline 
$\hat{\bar{D}}_{i}$ & 3 & 1 & 1/3 & $Q_{\bar{D}}$ \\ \hline 
$\hat{S}_{i}$ & 1 & 1 & 0 &$Q_{S}$ \\ \hline 
$\hat{N}^{c}_{i}$ & 1 & 1 & 0 & $Q_{N}$ \\ \hline 
$\hat{H}_{3}$ & 1 & 2 & $-1/2$ & $-Q_{2}$
\end{tabular}
\caption{The left-handed chiral superfields of the model with their quantum numbers under $SU(3)_{c}$, $SU(2)_{L}$, $U(1)_{Y}$ and $U(1)^{'}$. }
\end{center}
\end{table}

\bigskip

\begin{table}
\begin{center}
\caption{Nonuniversal boundary conditions for the large $s$ solution in the $\eta$ model. The first row for each soft supersymmetry breaking parameter gives its value at the high scale, while the second row gives its value at $M_{Z}$. The units are $GeV$.}
\begin{tabular}{ccccc}   
            &    1       &      2    &      3    &      4 \\ \hline   
$m_{1}^{2}$ & $(688)^{2}$& $(1037)^{2}$& $(814)^{2}$& $(582)^{2}$  \\
	    & $(102)^{2}$&$(167)^{2}$&$(339)^{2}$&$(313)^{2}$ \\ \hline
$m_{2}^{2}$ & $(1792)^{2}$&$ (3112)^{2}$& $(1922)^{2}$& $(1516)^{2}$  \\
	    & $(237)^{2}$&$(380)^{2}$&$(374)^{2}$&$(204)^{2}$ \\ \hline
$m_{S}^{2}$ & $(1525)^{2}$&$ (1922)^{2}$& $(1562)^{2}$& $(1068)^{2}$  \\
	    & $-(459)^{2}$&$-(685)^{2}$&$-(552)^{2}$&$-(588)^{2}$ \\ \hline
$m_{Q_{3}}^{2}$ & $(805)^{2}$& $(1620)^{2}$& $(722)^{2}$& $(482)^{2}$  \\
	    & $(138)^{2}$&$(74)^{2}$&$(60)^{2}$&$(48)^{2}$ \\ \hline
$m_{u_{3}^{c}}^{2}$ & $(1220)^{2}$& $(2299)^{2}$& $(1216)^{2}$& $(919)^{2}$  \\
	    & $(102)^{2}$&$(101)^{2}$&$(77)^{2}$&$(45)^{2}$ \\ \hline
$m_{D}^{2}$ & $(669)^{2}$& $(898)^{2}$& $(621)^{2}$& $(438)^{2}$  \\
	    & $(386)^{2}$&$(557)^{2}$&$(480)^{2}$&$(462)^{2}$ \\ \hline
$m_{\bar{D}}^{2}$ & $(669.5)^{2}$& $(898)^{2}$& $(621)^{2}$& $(438)^{2}$  \\
	    & $(362)^{2}$&$(489)^{2}$&$(476)^{2}$&$(449)^{2}$ \\ \hline
$(M_{4},~ M_{2})$ & $260$& $208$& $349$& $319$  \\
	    & $(36,~76)$&$(29,~61)$&$(51,~102)$&$(45,~93)$ \\ \hline
$(A_{s},~A_{Q})$ & $-2607$& $-1663$& $-2332$& $-3029$  \\
	    & $(-23,~193)$&$(-105,~158)$&$(-275,~269)$&$(-61,~238)$ \\ \hline
$M_{Z^{'}}$ & 649 & 974 & 786 & 835 \\ \hline
$\alpha_{Z-Z^{'}}$ & 0.003 & 0.005 & 0.006 & 0.007 \\ \hline
$\tan{\beta}$ & 7.06 & 3.35 & 1.92 & 5.78 \\ 
\end{tabular}
\end{center}
\end{table}

\begin{table}
\begin{center}
\caption{Same as Table III, but for the $\psi$ model.}
\begin{tabular}{ccccc}   
            &    1       &      2    &      3    &      4 \\ \hline   
$m_{1}^{2}$ & $(1087)^{2}$& $(1590)^{2}$& $(1383)^{2}$& $(848)^{2}$  \\
	    & $(489)^{2}$&$(644)^{2}$&$(354)^{2}$&$(237)^{2}$ \\ \hline
$m_{2}^{2}$ & $(3225)^{2}$&$ (4940)^{2}$& $(3651)^{2}$& $(2241)^{2}$  \\
	    & $(498)^{2}$&$(689)^{2}$&$(372)^{2}$&$(240)^{2}$ \\ \hline
$m_{S}^{2}$ & $(1609)^{2}$&$ (2317)^{2}$& $(3142)^{2}$& $(1857)^{2}$  \\
	    & $-(767)^{2}$&$-(1461)^{2}$&$-(796)^{2}$&$-(570)^{2}$ \\ \hline
$m_{Q_{3}}^{2}$ & $(1609)^{2}$& $(2498)^{2}$& $(1777)^{2}$& $(1013)^{2}$  \\
	    & $(96)^{2}$&$(216)^{2}$&$(370)^{2}$&$(250)^{2}$ \\ \hline
$m_{u_{3}^{c}}^{2}$ & $(2336)^{2}$& $(2572)^{2}$& $(1216)^{2}$& $(1536)^{2}$  \\
	    & $(114)^{2}$&$(204)^{2}$&$(185)^{2}$&$(185)^{2}$ \\ \hline
$m_{D}^{2}$ & $(61)^{2}$& $(884)^{2}$& $(1257)^{2}$& $(678)^{2}$  \\
	    & $(654)^{2}$&$(1142)^{2}$&$(712)^{2}$&$(498)^{2}$ \\ \hline
$m_{\bar{D}}^{2}$ & $(61)^{2}$& $(884)^{2}$& $(1257)^{2}$& $(678)^{2}$  \\
	    & $(532)^{2}$&$(1004)^{2}$&$(579)^{2}$&$(430)^{2}$ \\ \hline
$(M_{4},~ M_{2})$ & $375$& $544$& $471$& $363$  \\
	    & $(55,~110)$&$(80,~159)$&$(69,~138)$&$(53,~106)$ \\ \hline
$(A_{s},~A_{Q})$ & $-2281$& $-5081$& $-5106$& $-3625$  \\
	    & $(-331,~293)$&$(-104,~407)$&$(58,~347)$&$(-19,~269)$ \\ \hline
$M_{Z^{'}}$ & 1087 & 1561 & 1120 & 809 \\ \hline
$\alpha_{Z-Z^{'}}$ & 0.001 & 0.0005 & 0.0007 & 0.002 \\ \hline
$\tan{\beta}$ & 0.97 & 0.74 & 0.80 & 0.70 \\ 
\end{tabular}
\end{center}
\end{table}

\bigskip
\begin{figure}
\centerline{
\hbox{
\epsfxsize=2.8truein
\epsfbox[70 32 545 740]{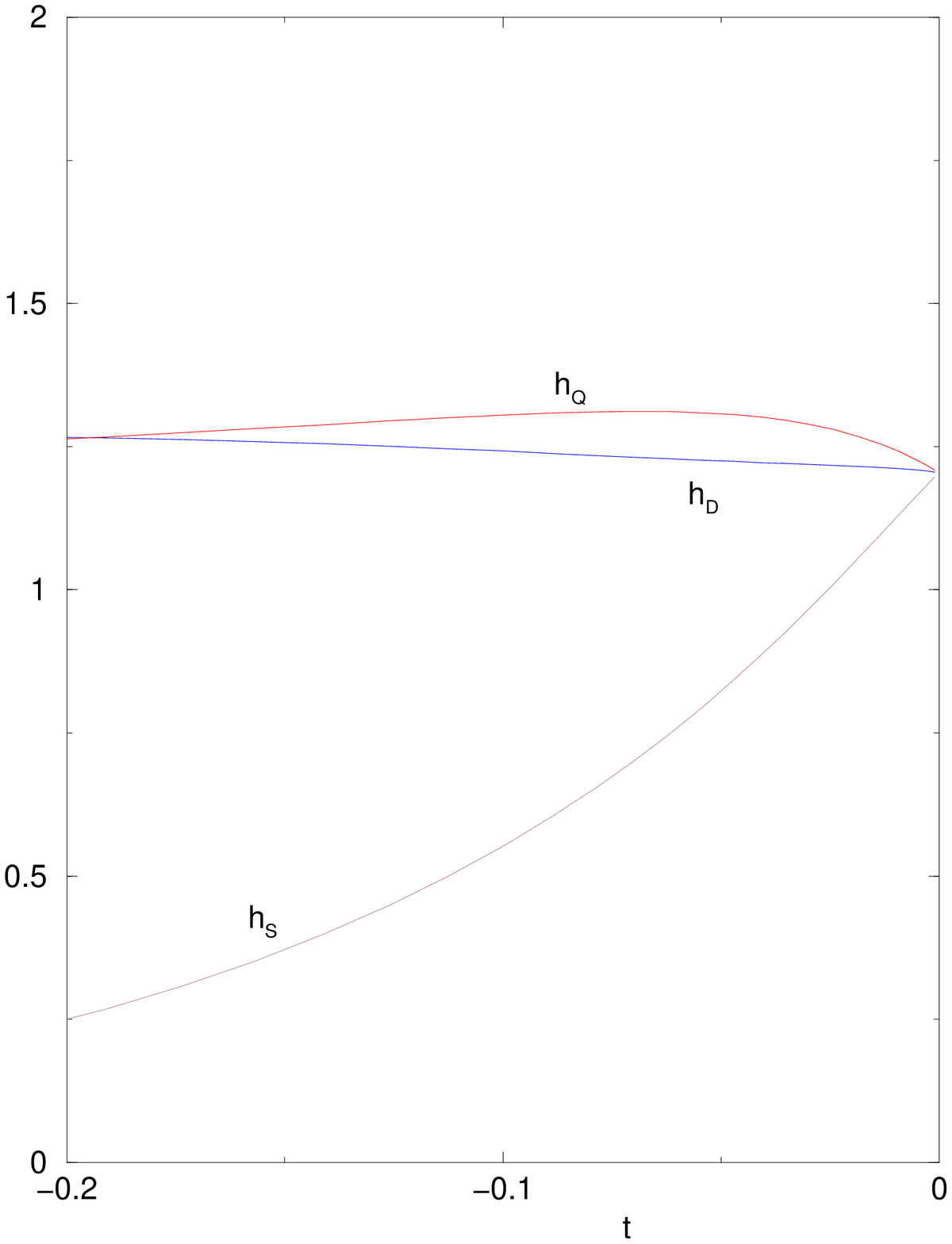}
\hskip 0.25truein
\epsfxsize=2.8truein
\epsfbox[70 32 545 740]{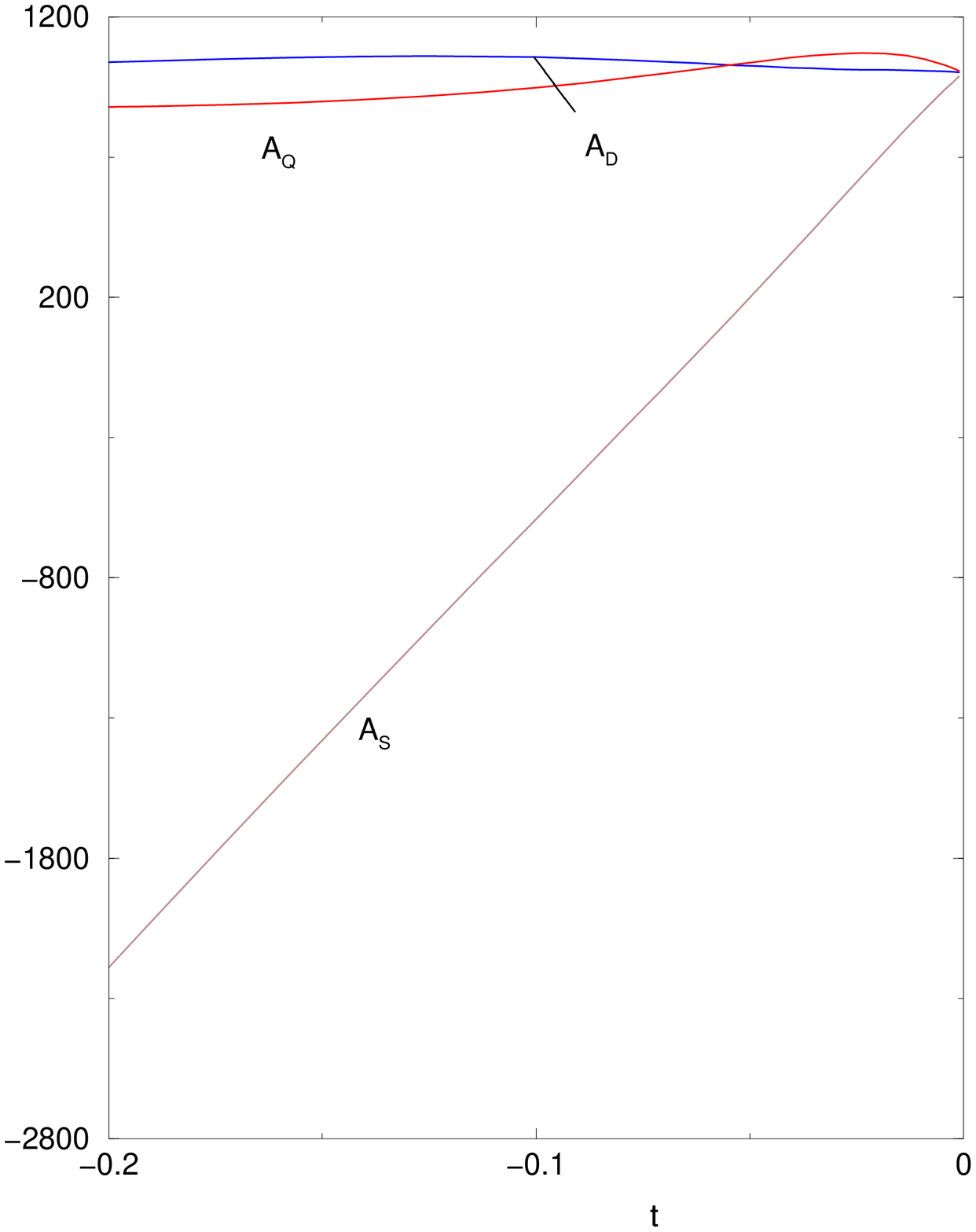}
}
}
\caption{
Running of the Yukawa and trilinear couplings, with universal boundary conditions and $c_{A}=c_{1/2}=1.0$. 
}
\vskip 0.1truein
\centerline{
\hbox{
\epsfxsize=2.8truein
\epsfbox[70 32 545 740]{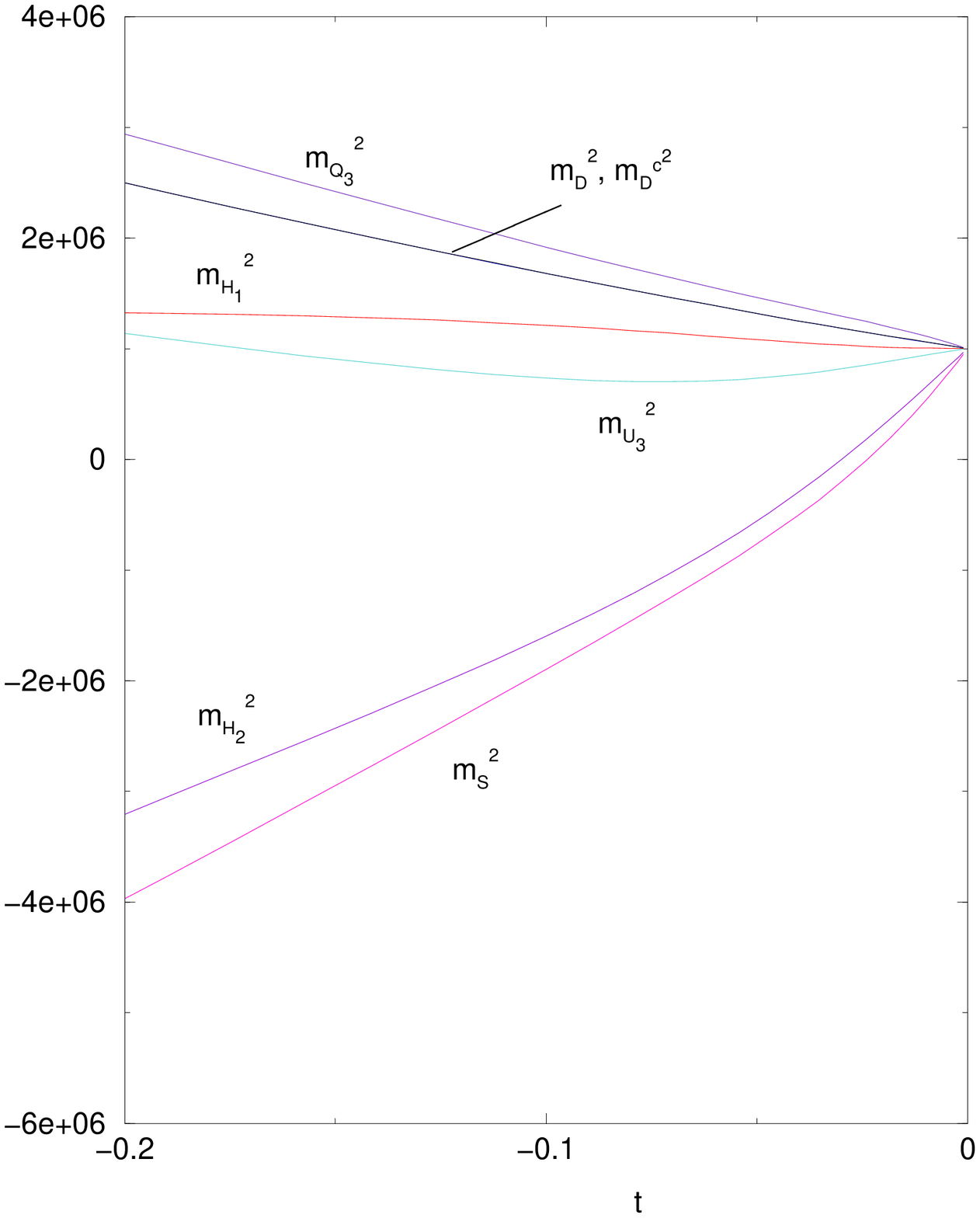}
}
}
\caption{
Running of mass squares, with universal boundary conditions and $c_{A}=c_{1/2}=1.0$.
}
\end{figure} 
 
\begin{figure}
\centerline{
\hbox{
\epsfxsize=2.8truein
\epsfbox[70 32 545 740]{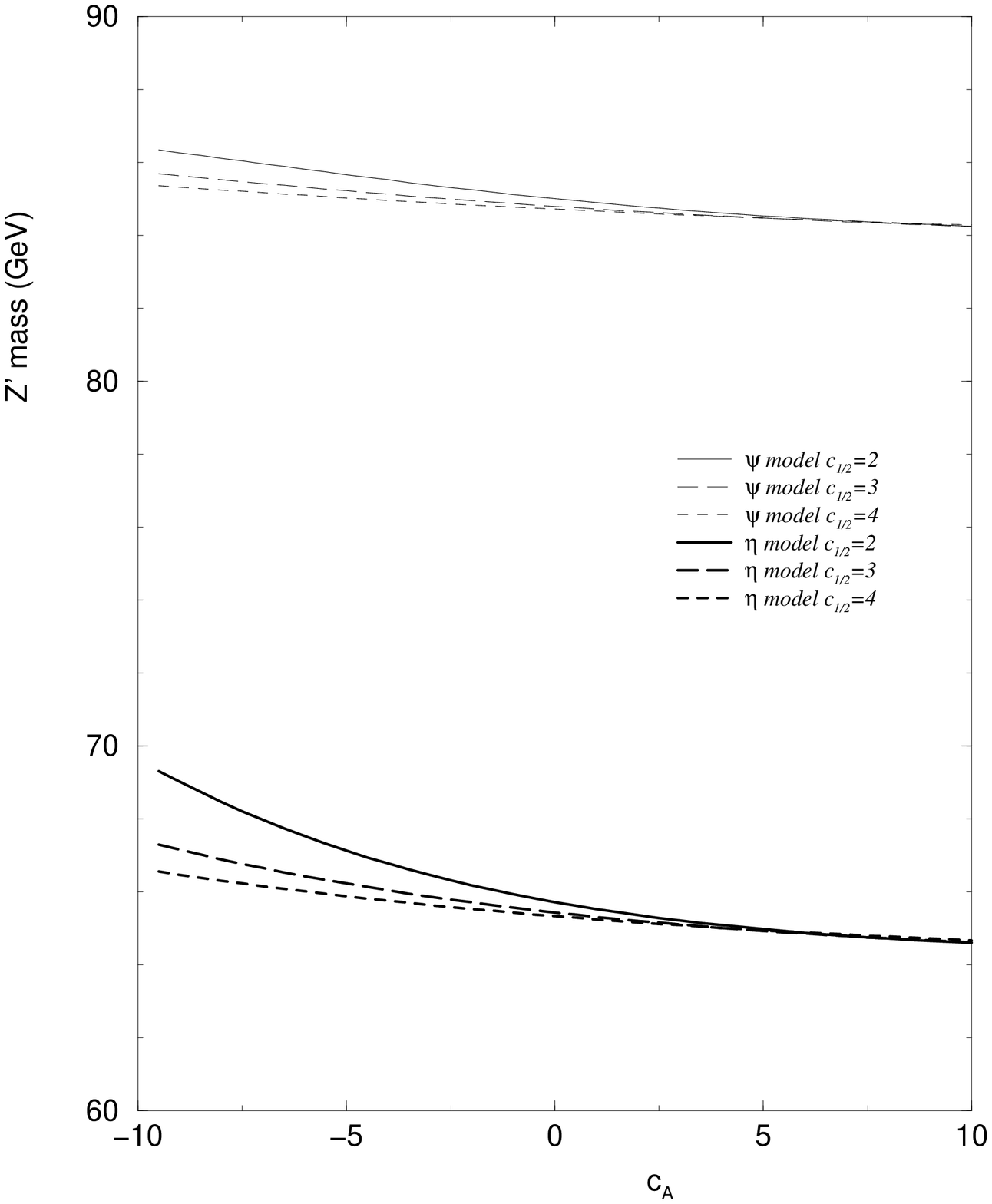}
}
}
\caption{
The $Z^{'}$ mass as a function of $c_{A}$ for various $c_{1/2}$, with universal boundary conditions. 
}
\vskip 0.1truein
\centerline{
\hbox{
\epsfxsize=2.8truein
\epsfbox[70 32 545 740]{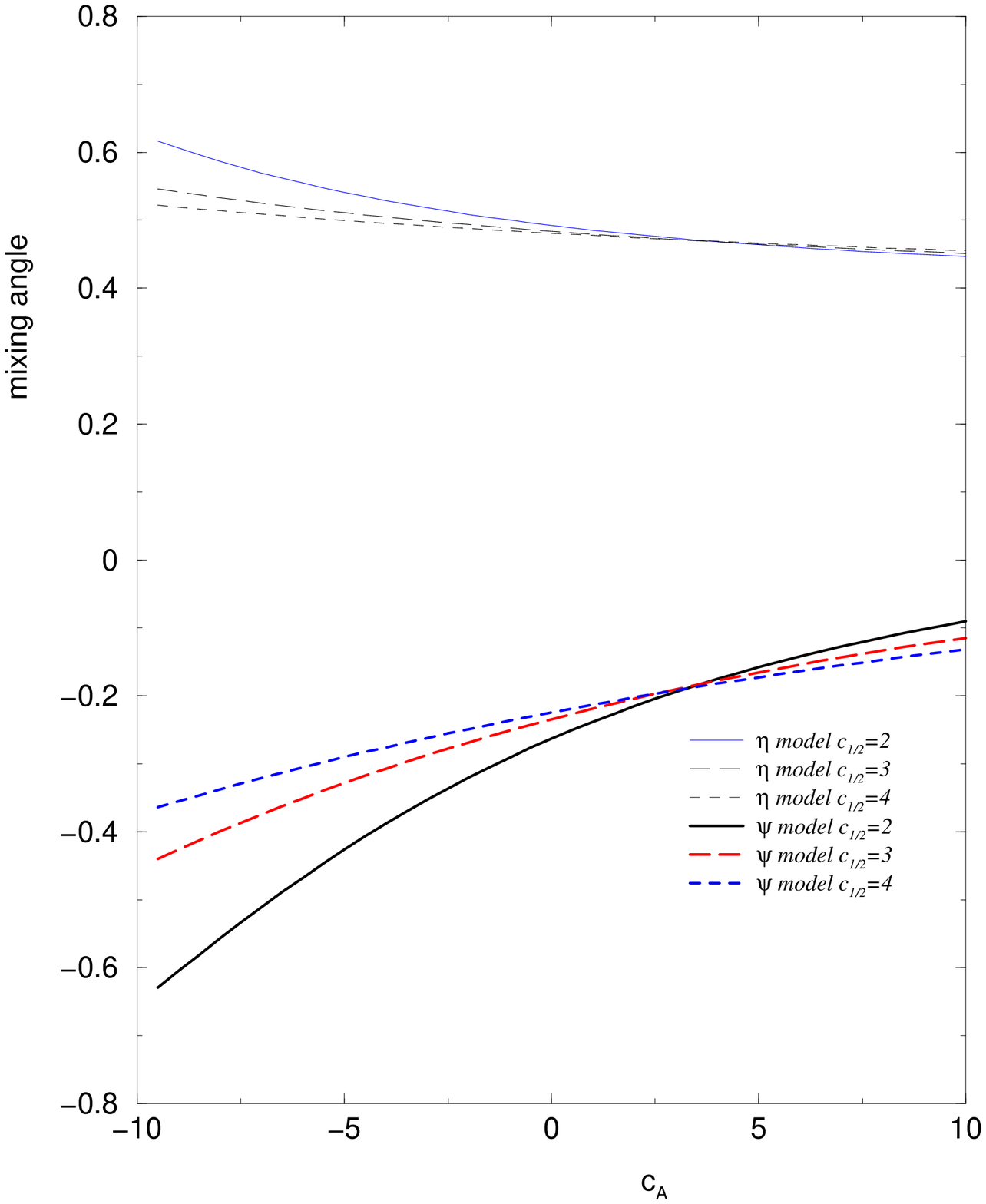}
}
}
\caption{
The $Z^{'}-Z$ mixing angle $\alpha$ as a function of $c_{A}$ for various $c_{1/2}$, with universal boundary conditions. 
}
\end{figure} 

\begin{figure}
\centerline{
\hbox{
\epsfxsize=4.4truein
\epsfbox[55 32 525 706]{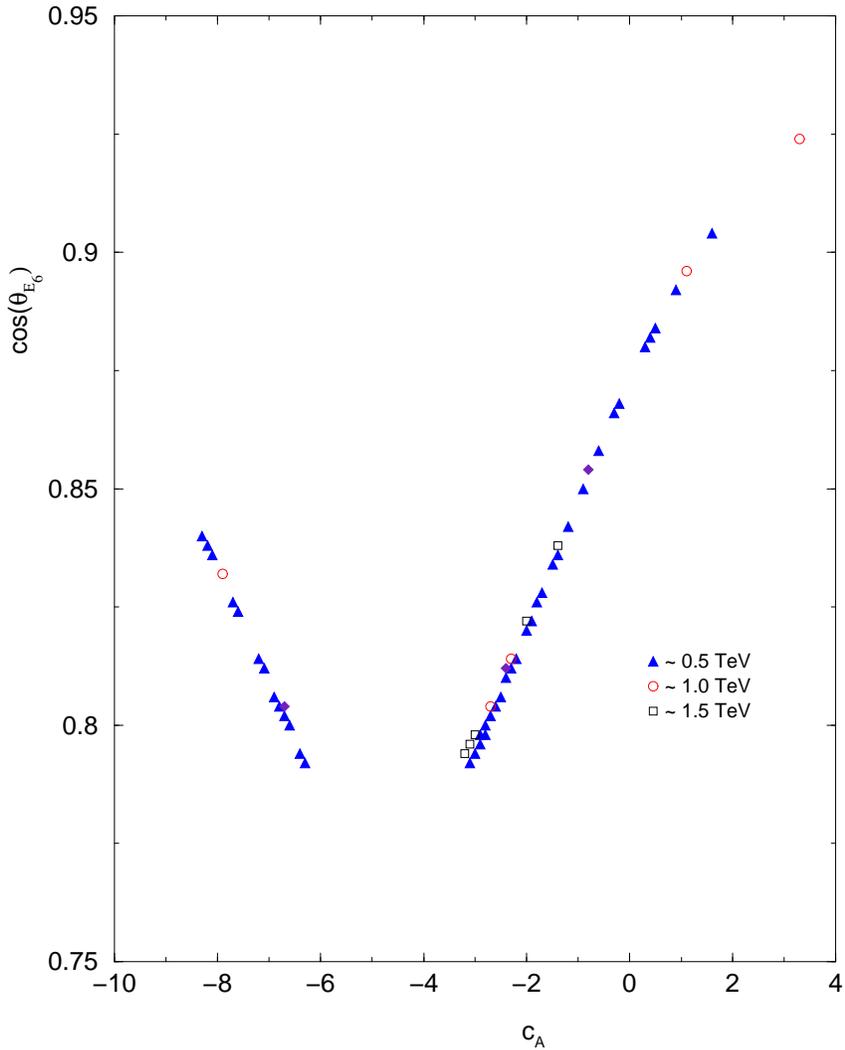}
}
}
\caption{
Contour plot of the ($c_{A}$, $\cos{\theta_{E_{6}}}$) parameter ranges for
$M_{Z^{'}}>500~GeV$, $c_{1/2}=0.4$, where $c_{A}$, $c_{1/2}$ are the
dimensionless parameters for the trilinear couplings $A_{i}$ and the gaugino
masss; $\theta_{E_{6}}$ is the mixing angle between the $U(1)_{\chi}$
and $U(1)_{\psi}$ of $E_{6}$ .     
}
\end{figure}

\end{document}